\documentclass[journal]{IEEEtran}
\usepackage[utf8]{inputenc}
\usepackage{amsmath,amssymb}
\usepackage{booktabs,tabularx,multirow,array}
\usepackage{graphicx}
\usepackage{xcolor}
\usepackage{xurl}

\usepackage[hidelinks]{hyperref}
\usepackage{cite}
\newcolumntype{Y}{>{\raggedright\arraybackslash}X}

\begin{document}
\nocite{nerc_ir2025,wharton2026pjm,nerc_l3alert,ercot_nogrr282,aemc2026final,energinet2024tr343,eirgrid2026mpid345,caiso_straw,spp_rt2026,cigre2026cse41,asme_vv40,s1_surrogate,yogarathnam2026bess,cheerla2026emtrms,cheerla2026lvrt,nerc_trd,pnnl_emt,sun2022p1,nerc_wp_disturbance,ieeesa_ic25004,esig2026behav,ahmed2021,ahmed2022,zechevikj2026,yazdanpanah2026partial,mishra2026discerning,sun2022p2,s3_147hz,biswas2026,azizi2026,shen2025,barati2026,chaudhary2026,chaudhary2026b,jimenezortega2026,afzaal2025,zhu2023,sanati2026,kosterev2026perc1,soco2026tcll,albustami2026attribution,epri2026rt,spp_ops2026,liu2025,ercot2026puct,cru2026mpid345,statnett2018,alapera2019,jenkins2024wecc,ieso2026lcl,esig2026perf,appa2026nerc,soni2026,neso2025gcdf,entsoe2025cnc,meier2026nercclo,conto2025,rashwan2026,thomas2026,meier2026,hussein2026,norouzi2026dualpq,wang2026solvrt,acer2025dc,xie2026eenergy,eirgrid2026framework,syed2026esig,soco2025ibrmodel,nerc_mod026,milanovic2013,arif2018,sun2019pfc,odunlami2026perc1,sleiman2025cmld,kosterev2008,nerc2016dlm,wecc_cmld_spec,entergy2026,nerc_questionnaire,powerworld_ldev1,javadi2026review,ercot_pscad,samanta2024emt,zamani2026esig,jimenezruiz2026,arifin2026,gyang2026tpwrs,ross2026,lu2026,lu2026pattern,sreejith2026,silva2026,samanta2024,ahmad2025,sun2021,sun2022,tang2026,ramasubramanian2025,bindal2024npsc,reyes2024,ali2026steady,miah2026,hossain2025,barati2025,badakhshan2026,zhabelova2015,shamseldein2026,zubi2024,aghadinuno2026,harzevili2026,wang2020ddqn,bu2020edrl,qiu2026bayes,zhu2026diffusion,xiao2023neuralode,shen2024neudye,lin2026rlsr,dorji2026pdkan,qi2022pmu,mitrentsis2021adn,paidi2020ocl,leinakse2021dfr,lin2024taiwan,avila2020cmld,lin2026identifiability,woo2026reinsertion,zhang2024lstmdcdc,varshney2025sysid,lu2025uniqueness,transtrum2025tailoring,cicilio2020metrics,elsayed2026dataset,govind2023hpc,saroare2026dcclm,shenoy2026switched,ruan2025ssr,acero2026processes,ventura2026pinn,guruwacharya2024phil,subedi2025overview,nasa7009,choi2026kiee,alenius2020,alkez2020,khomarudin2025,azizi2026b,lyu2026,mohammadi2026adapen,shamseldein2026b,kundu2026,azizi2025,seo2026,gu2026,ilic2026,elliott2026sandia,ghosh2025,shamseldein2027,tao2025,kim2025,sun2025,maheshwari2026,viththarachchige2026,ghazavi2026pem,gilani2026energies,shrestha2026esm,shihab2027apen,singh2025aireview,gyang2026ojies,dhakal2026review,xie2026procieee,chen2026nexus,mughees2026review,song2026ojpel,ginzburg2026energies,seshmasetti2025isgt,parikh2025,wang2026,chaudhary2026semimarkov,ko2026wide}

\title{Which Data-Center Load Models Can Be Trusted?\\ An Evidence-Graded Review of Dynamic Modeling and Validation for Large Computational Loads}

\author{Dileep~Cheerla,~\IEEEmembership{Member,~IEEE,}
and~Amirthagunaraj~Yogarathnam,~\IEEEmembership{Member,~IEEE}
\thanks{D. Cheerla is an independent researcher, Dallas, TX, USA (e-mail: dileepreddy.cheerla@gmail.com; ORCID 0009-0004-6581-1899). He is the corresponding author.}
\thanks{A. Yogarathnam is an independent researcher, Dallas, TX, USA (e-mail: amirthegunaraj@gmail.com; ORCID 0000-0003-3260-0492).}
\thanks{This work has been submitted to the IEEE for possible publication. Copyright may be transferred without notice, after which this version may no longer be accessible.}
\thanks{This work received no external funding.}}

\markboth{Cheerla \MakeLowercase{\textit{et al.}}: Which Data-Center Load Models Can Be Trusted? (preprint)}{Cheerla \MakeLowercase{\textit{et al.}}: Which Data-Center Load Models Can Be Trusted? (preprint)}

\maketitle

\begin{abstract}
Grid operators now require dynamic models of data centers and other large computational loads, and new ride-through rules assume that these models predict how facilities respond to disturbances. This review asks what evidence supports that assumption. We coded 160 documents (from 1{,}424 de-duplicated records plus 65 added) by the strongest reference object each presents, from grid-side records to simulation only, and separately 70 data-driven model studies. Across 19 field-evidence entries, 15 requirement documents and ten model classes, we find two comparisons of a standard library model with measured data-center events: with default parameters both reported a mismatch, though neither stated an acceptance criterion, and per-event refitting reduced the error but was never tested on an unseen event. Among the sources reviewed, we found no successful held-out validation of a frozen, facility-specific data-center model against an independent measured disturbance, and 25 of the 33 model--phenomenon combinations with any evidence involve no electrical measurement. The features that distinguish the new rules (memory of repeated dips, recovery within 0.5--3\,s, proportional power reduction during the sag, overvoltage response) are those current phasor models cannot represent or were never tested against. No operator procedure examined requires comparing the submitted model with a measured facility response against a predefined acceptance criterion, as generator standards do, and proposed mitigations have mostly been evaluated on unvalidated facility models. We propose context-of-use gates that separate calibration from validation, release labels for model claims, a minimum validation protocol with acceptance criteria, and a research agenda.
\end{abstract}


\section{Introduction}\label{sec:intro}
\IEEEPARstart{L}{arge} computational loads have become a reliability concern within three years. On 10 July 2024 a single 230\,kV fault with repeated reclosing in Northern Virginia caused about 1{,}500\,MW of data-center load to transfer to backup power and stay off the grid for hours~\cite{nerc_ir2025}. Similar customer-initiated load reductions have since been recorded in several regions, the largest of them about 3{,}800\,MW on 22 July 2026~\cite{wharton2026pjm}. Unlike a conventional load, a data center responds to a voltage dip through the protection and control logic of its uninterruptible power supplies and power-supply units, and that logic is set by the owner, often not disclosed, and changed with firmware.

\begin{figure*}[!t]
\centering
\includegraphics[width=\textwidth]{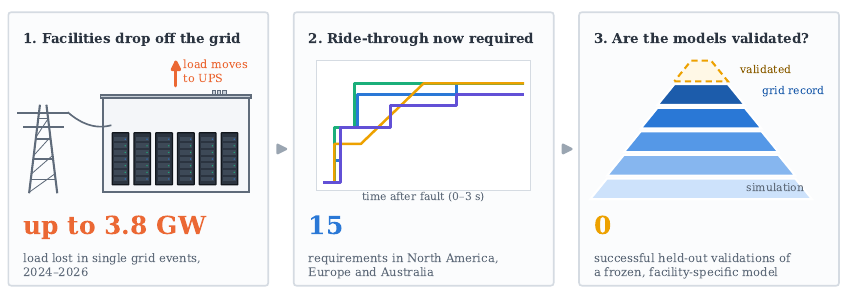}
\caption{The question this review asks. Large computational loads have dropped up to 3.8\,GW in single grid events (Section~\ref{sec:events}); operators have answered with ride-through requirements that are demonstrated with dynamic models; yet the measured evidence about those models is thin: two independent comparisons of a generic library model with measured events both reported a mismatch, and we found no successful held-out validation of a frozen, facility-specific model.}
\label{fig:ga}
\end{figure*}

Regulators have responded quickly. NERC issued a Level~3 alert that sets a minimum dynamic model for large loads~\cite{nerc_l3alert}, and operators in North America, Europe and Australia have adopted or proposed ride-through requirements with explicit voltage, duration and recovery limits; Australia's took effect on 12 November 2026~\cite{ercot_nogrr282,caiso_straw,spp_rt2026,eirgrid2026mpid345,energinet2024tr343,aemc2026final}. Interconnection studies now use dynamic models to show that facilities will meet these requirements, and to size the mitigations that make them do so. Yet a submission to the Australian rule change argued that models validated to the standard of generator models ``may be infeasible'' for data centers~\cite{aemc2026final}, and CIGRE's large-load task force calls for routine validation ``similar to generator model validation''~\cite{cigre2026cse41}.

The 20 recent reviews compared in Table~\ref{tab:positioning} describe architectures, models, requirements and flexibility. None of them asks the question on which these studies depend: which model claims are supported by measured evidence, and for what decisions? Model credibility frameworks from other fields~\cite{asme_vv40} distinguish calibration from validation and tie the evidence required to the decision a model supports. This review applies that discipline to dynamic models of data centers. Its contributions are:
\begin{itemize}
\item an evidence-graded audit of 160 documents, coding each by the strongest reference object it presents and by whether the object validated is the object claimed (Sections~\ref{sec:method}--\ref{sec:matrix});
\item a catalog of 19 field-evidence entries (single events, event series, operating records and demonstrations), a comparison of fifteen ride-through requirements, and an analysis of the evidence each operator accepts that a model represents the facility (Section~\ref{sec:events});
\item a context-of-use framework with gates G0--G7 and release labels for model claims (Section~\ref{sec:framework}), applied to eleven worked cases (Section~\ref{sec:cases});
\item a requirement-to-model mapping, an evidence audit of proposed mitigations, and a minimum validation protocol (Sections~\ref{sec:mapping}--\ref{sec:protocol});
\item a positioning against 20 related reviews and a research agenda (Sections~\ref{sec:positioning}--\ref{sec:agenda}).
\end{itemize}
The central finding is that, among the sources reviewed, no frozen, facility-specific data-center model has been successfully validated against a held-out measured disturbance, while the requirements those models are used to demonstrate are already in force, and none of the operator procedures examined requires an independent comparison of the submitted model with a measured facility response against a predefined acceptance criterion.

\section{Review Method and Evidence Grading}\label{sec:method}
\begin{figure}[!t]
\centering
\includegraphics[width=0.92\columnwidth]{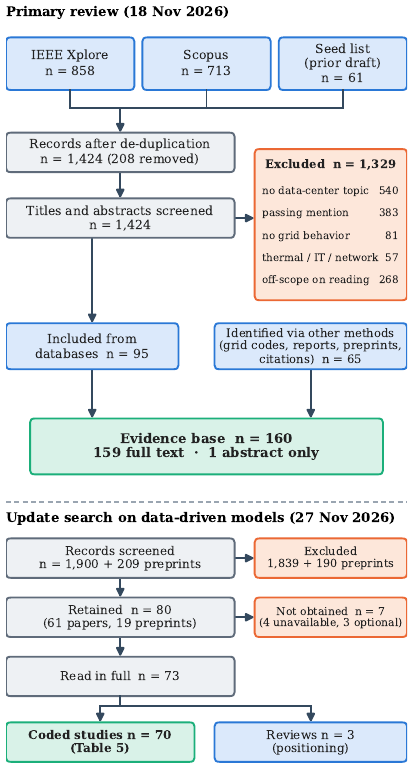}
\caption{Identification and screening of sources: primary review (top) and update search on data-driven models (bottom).}
\label{fig:prisma}
\end{figure}

This is a critical evidence review rather than a full systematic review: its aim is to establish what the available evidence can support, not to enumerate every publication. The search and screening are nevertheless documented so that they can be repeated (Fig.~\ref{fig:prisma}).

\emph{Search.} On 18 November 2026 we searched IEEE Xplore (858 records) and Scopus (713 records, from 2015) with queries combining data-center, large-load and computational-load terms with model, ride-through, electromagnetic-transient, voltage-sag, oscillation and validation terms, and added 61 records from an earlier working draft. The Scopus search was limited to publications from 2015 onward, so older laboratory studies of the sag immunity of computer power supplies are under-represented in the device-level rows. After de-duplication by DOI and normalized title, 1{,}424 records remained.

\emph{Screening.} Records were included if they concerned the grid-facing electrical behavior, modeling, validation or requirements of data centers or other large computational loads. A keyword pre-screen removed records with no data-center subject, with only a passing mention, with no grid-facing behavior, or about thermal, IT or network topics; the remaining titles and abstracts were read. This left 95 records. We added 65 documents not indexed in the two databases: operator and regulator reports and grid codes from North America, Europe and Australia, working-group presentations, standards-body reports, preprints, and sources found by following citations. Of the resulting 160 documents, 159 were read in full and one, a neural-surrogate screening study~\cite{s1_surrogate}, only as an abstract (it sets no matrix cell); a presentation cited by another source could not be obtained and is not counted, and duplicates and two received documents outside the scope were removed. Every document, its type, origin, access and coding are listed in a source register available from the corresponding author on reasonable request.

\emph{Coding.} Each source was coded for the strongest reference object it presents (tiers below), for the model class and phenomena it addresses, and for whether the object it validated is the object of its claims. Of the 160 documents, 75 studies were coded into at least one model-class and phenomenon pair, 30 studies were recorded as outside the matrix's scope with a reason (for example, frequency-support services, faults internal to the facility, or steady-state power-quality models), and 55 reviews, requirement documents and practice documents were used for context but not coded into the matrix. Practice documents and event reports that themselves report observations, parameters or simulations of a model class (for example the NERC technical reference, the NERC draft disturbance white paper, a WECC presentation and the NERC questionnaire) were coded as studies for that content and are among the 75. Each matrix cell is then derived as the highest-provenance code over all coded sources. Coding was performed by a single reviewer; this is a limitation, and the full coding is available from the corresponding author on reasonable request. One included study, a mitigation study, is co-authored by both authors~\cite{yogarathnam2026bess}; it was coded by the same rules as every other source. Two of the first author's studies of inverter-based resources are cited for context only and are not part of the reviewed corpus~\cite{cheerla2026emtrms,cheerla2026lvrt}.

\emph{Update search on data-driven models.} The original searches did not target data-driven and AI-based load models, so on 27 November 2026 we ran an update search: three Scopus searches combining machine-learning and data-driven terms with load-model terms, with data-center and AI-workload terms, and with converter, power-supply and UPS terms, together with three targeted searches run in both IEEE Xplore and Scopus. After removing duplicates and the 12 records already in the corpus, 1{,}900 records and 209 preprints were screened by the criteria above; 80 documents were retained, of which 73 could be obtained and read in full (four were not available, and three low-priority preprints on load shaping, chip-level GPU voltage effects and language-model agents for inverter modeling were not pursued because none concerns load-model validation). Of the 73, 70 are studies coded in Section~\ref{sec:datadriven} and three are reviews used for positioning (Section~\ref{sec:positioning}). They are coded separately in Section~\ref{sec:datadriven} and leave the 160-document corpus, the evidence matrix and every count elsewhere in this review unchanged, for the reason given there.

The reference-object tiers are:
\begin{itemize}
\item \textbf{T1} measured grid-side record at or beyond the point of interconnection (POI): PMU, DFR, relay oscillography;
\item \textbf{T2} measurement inside the facility (switchboard, rack, relay log);
\item \textbf{T2w} IT-workload telemetry (CPU/GPU utilization, scheduler logs) used as model input;
\item \textbf{T3} laboratory, device or hardware-emulator test;
\item \textbf{T4} parameters from OEM documentation or operator questionnaires;
\item \textbf{T5} simulation only, including model-to-model comparison.
\end{itemize}
Tiers describe provenance, not quality: a T5 study can be rigorous, and a T1 record can support only a narrow claim.

\section{How a Data Center Looks From the Grid}\label{sec:physics}
\begin{figure*}[!t]
\centering
\includegraphics[width=0.95\textwidth]{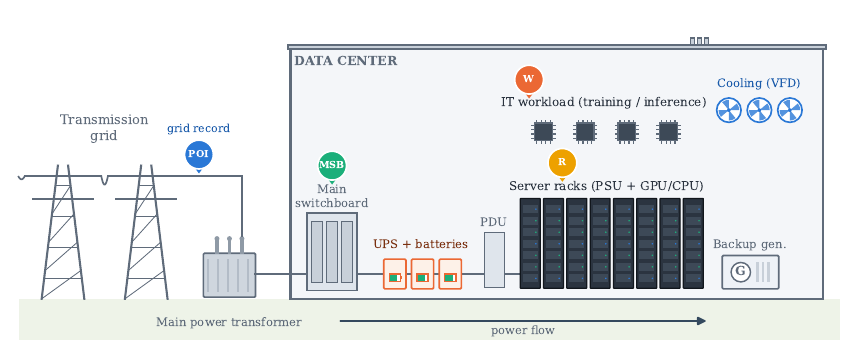}
\caption{Simplified power path of a large data center and the four measurement boundaries used in this review: IT-workload telemetry (W), rack or server terminals (R), main switchboard (MSB), and the point of interconnection (POI). Evidence gathered at one boundary does not by itself establish behavior at another.}
\label{fig:path}
\end{figure*}

A large data center reaches the transmission system through one or more main power transformers, medium-voltage collection and unit substations, and a low-voltage distribution chain whose dominant element is the double-conversion uninterruptible power supply (UPS)~\cite{nerc_trd,pnnl_emt}. Downstream of the UPS, power-distribution units feed server power supplies whose active power-factor-correction (PFC) stages regulate a DC bus, so that the IT load seen from the grid behaves approximately as constant power during voltage dips~\cite{nerc_trd,sun2022p1}. Cooling is supplied increasingly through variable-frequency drives or electronically commutated motors and is therefore also electronically interfaced. On-site batteries and backup generators complete the picture. Fig.~\ref{fig:path} marks the four places at which this chain is observed in the literature reviewed here.

\subsection{Behavior during voltage disturbances}
\begin{figure}[!t]
\centering
\includegraphics[width=\columnwidth]{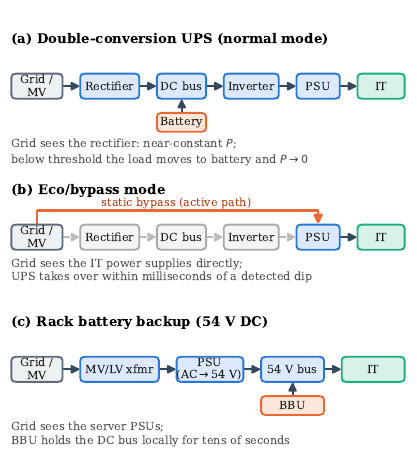}
\caption{Three common power-chain configurations and what the grid sees during a voltage dip~\cite{esig2026behav,ieeesa_ic25004}. The same facility can move between (a) and (b) with its operating mode, so a model parameterized in one mode does not describe the other.}
\label{fig:ups}
\end{figure}

\begin{figure}[!t]
\centering
\includegraphics[width=\columnwidth]{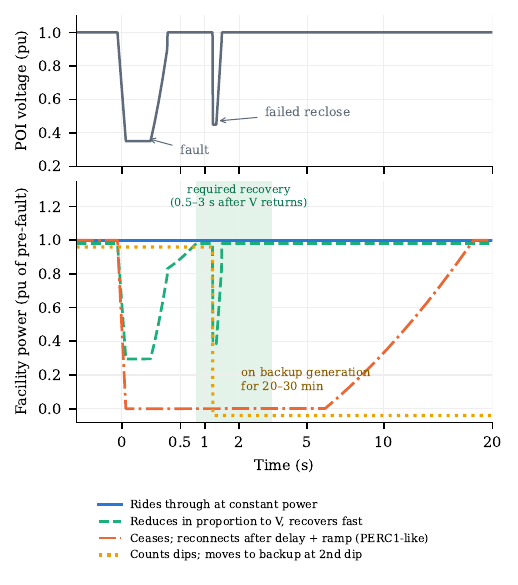}
\caption{Illustrative (not measured) responses of a facility to a fault followed by a failed reclose. Each behavior has been reported in field events~\cite{nerc_wp_disturbance,wharton2026pjm}; requirements now expect recovery within 0.5--3\,s (shaded), which the delay-and-ramp behavior of typical PERC1 settings~\cite{kosterev2026perc1} and a transfer to backup generation~\cite{esig2026behav} cannot meet. Curves are drawn with a 0.7\,pu ceasing threshold, a 6\,s reconnection delay and a 12\,s ramp for the PERC1-like case. Time axis is square-root scaled.}
\label{fig:dip}
\end{figure}

Event records show that facilities that ride through a dip draw essentially constant power during it, so their current rises as voltage falls~\cite{nerc_trd}. In one ERCOT event the current of two cryptocurrency-mining loads rose to 150--160\,\% of its pre-fault value~\cite{nerc_wp_disturbance}. When a dip is deep, long or repeated enough, protection or UPS control transfers the IT load to the UPS battery and, for longer events, to backup generation. Ride-through thresholds are rooted in the ITIC (formerly CBEMA) curve, and planning guidance suggests assuming that a dip below 0.7\,pu lasting 20\,ms or more causes a transfer for the rest of a simulation when settings are unknown~\cite{nerc_trd}. How a facility reaches its stored energy depends on its UPS architecture: a double-conversion UPS isolates the IT load continuously; an eco-mode (high-efficiency bypass) UPS re-assumes the load within a few milliseconds of a detected disturbance; and in-rack battery backup units hold up the DC bus locally for tens of seconds~\cite{ieeesa_ic25004}. Measured IT-equipment dropout thresholds cluster around 0.5--0.7\,pu depending on sag duration, and UPS logic is tuned to protect the IT load first, which can conflict with grid ride-through expectations~\cite{ieeesa_ic25004}. A facility that has transferred to backup generation typically cannot reconnect until the generators' minimum run time of about 20--30\,min has elapsed~\cite{esig2026behav}. Internal faults matter too: cable faults inside the facility's own distribution can produce dips that violate the ITIC tolerance of server power supplies even when the grid is healthy~\cite{ahmed2021,ahmed2022}, and the dead time of transmission autoreclosing can hold the bus below UPS thresholds long enough to disconnect a facility that would have ridden through the fault itself~\cite{zechevikj2026}. Fig.~\ref{fig:ups} shows what the grid sees in each case. The UPS architecture and its operating mode are therefore part of the context in which any model claim holds (Section~\ref{sec:framework}). Fig.~\ref{fig:dip} illustrates the four behaviors that a model may have to represent. Two further features make this behavior hard to model. First, many facilities count voltage dips within a time window and transfer only after the $n$-th strike~\cite{nerc_wp_disturbance}. Second, reconnection is not immediate: reported returns range from under a minute to hours~\cite{nerc_wp_disturbance}. Unbalanced faults add a third feature: reductions concentrate on the faulted phases and a facility may keep operating unbalanced after clearing~\cite{nerc_wp_disturbance,yazdanpanah2026partial}.

\subsection{Workload-driven variation and resonance}
AI training synchronizes thousands of accelerators, so facility demand can swing periodically. Dominion's year-long synchrophasor record showed a $\sim$0.40\,Hz duty cycle that had first been mistaken for a stability problem~\cite{mishra2026discerning}. At higher frequencies, interactions between power-supply input impedance and the facility network produced 49/71\,Hz components in a Meta facility~\cite{sun2022p1,sun2022p2}, and Dominion observed a 14.7--14.8\,Hz oscillation emerging from a data-center-rich area that lasted about two hours; synchrophasor analysis tracked a 10--11\,Hz mode losing damping into a limit cycle, and the facility owner confirmed that the input stage of its UPS units caused it~\cite{s3_147hz}. Although a single GPU can change its power within microseconds, the input current of its power supply is held for one to two cycles, so facility-level ramps are limited to roughly 50\,ms~\cite{esig2026behav}. The simulation literature on these phenomena is now large: forced-oscillation risk screening~\cite{biswas2026}, sub-synchronous oscillation driven by workload or by clustered UPS converters~\cite{azizi2026,shen2025,barati2026}, spatially correlated load dynamics~\cite{chaudhary2026,chaudhary2026b}, flicker in weak grids~\cite{jimenezortega2026}, supraharmonic emission beyond the reach of PMUs~\cite{afzaal2025}, impedance aggregation of heterogeneous PFC converters~\cite{zhu2023}, and power-swing encroachment on distance-relay characteristics~\cite{sanati2026}. Almost all of it uses synthetic or telemetry-derived load profiles; one study notes explicitly that real facility signatures cannot be confirmed without proprietary facility-level data~\cite{chaudhary2026b}. Operators have begun to set measurable limits instead: Southern Company caps the spectral content of a facility's active power at the point of connection at 10\,MW between 0.1 and 5\,Hz and 2.5\,MW between 5 and 55\,Hz, evaluated in 0.1\,Hz bins~\cite{soco2026tcll}.

\subsection{What each measurement boundary can show}
The boundary at which evidence is collected limits the claim it can support (Fig.~\ref{fig:path}). Workload telemetry (W), such as GPU utilization traces, describes what the IT equipment was asked to do, not how the power chain and protection translated it into POI current. Rack and switchboard measurements (R, MSB) observe internal electrical behavior but not the grid-facing aggregate. POI recorders observe the aggregate but cannot always tell why it changed: a workload reduction and a transfer to backup supply can produce identical POI voltage and current in a quasi-static constant-power model~\cite{albustami2026attribution}. Even the voltage itself depends on where and how it is defined: utilities may assess compliance on positive-sequence or on the most severe phase voltage, which differ substantially during unbalanced faults, and equipment inside the facility can see more severe voltages than the POI measurement implies because of transformer connections and internal impedances~\cite{epri2026rt}. Instrument bandwidth matters as well. PMUs reporting 30 or 60 frames per second cannot resolve the upper sub-synchronous range; NERC notes that this rate is not sufficient for the upper sub-synchronous range and that higher reporting rates, such as 120 or 240 per second, would improve accuracy~\cite{nerc_trd}, and SPP plans to require higher-resolution recorders for that reason~\cite{spp_ops2026}. Wide-area frequency monitoring can detect load-loss events from voltage signatures~\cite{liu2025}, but detection is not the same as recording the facility's response at the bandwidth a model comparison needs.

\section{Field Evidence and Emerging Requirements}\label{sec:events}

\subsection{Documented disturbance events}
\begin{figure*}[!t]
\centering
\includegraphics[width=\textwidth]{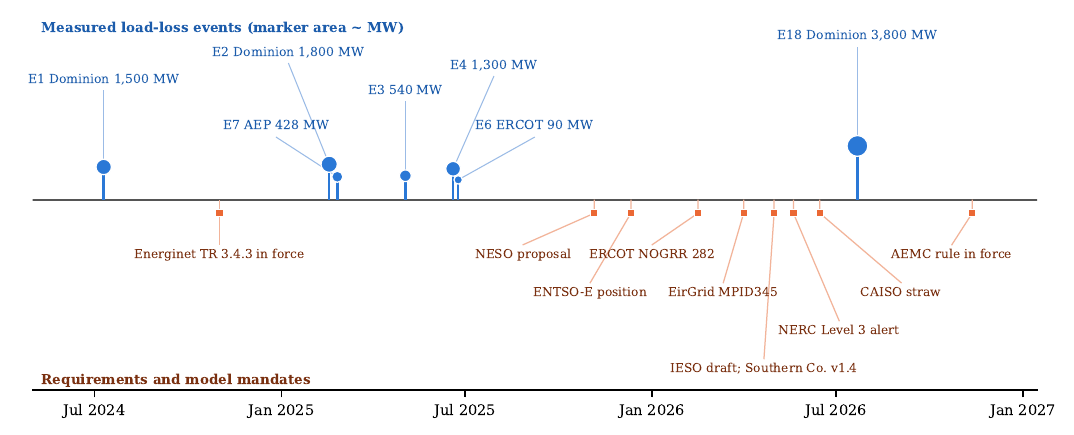}
\caption{Measured load-loss events since 2024 (Table~\ref{tab:events}; Irish events from 2022 and the multi-year ERCOT series E5 are omitted for clarity; requirements are placed at the date of the document cited in Table~\ref{tab:reqs}, which may be a proposal, approval or entry-into-force date) and the requirements and model mandates that followed (Table~\ref{tab:reqs})~\cite{nerc_ir2025,nerc_wp_disturbance,wharton2026pjm,energinet2024tr343,neso2025gcdf,entsoe2025cnc,ercot_nogrr282,eirgrid2026mpid345,ieso2026lcl,soco2026tcll,nerc_l3alert,caiso_straw,aemc2026final}.}
\label{fig:timeline}
\end{figure*}

Table~\ref{tab:events} lists the published field observations of the responses of computational loads, with event type and measurement boundary, and Fig.~\ref{fig:timeline} places the largest beside the requirements that followed. In Northern Virginia, four events between July 2024 and June 2025 produced customer-initiated load reductions (CILR) of about 540--1{,}800\,MW~\cite{nerc_ir2025,nerc_wp_disturbance}, and on 22 July 2026 about 3{,}800\,MW transferred to backup supply after a single, correctly cleared 230\,kV fault, the largest such event PJM has experienced~\cite{wharton2026pjm}. ERCOT has analysed 32 ride-through events of 100\,MW or more since January 2023, 28 of them involving cryptocurrency mining and none so far a traditional data center~\cite{nerc_wp_disturbance}. EirGrid reports reductions of up to 52\,\% of the connected data-center load, with RoCoF of about 0.26\,Hz/s~\cite{nerc_wp_disturbance}.

\begin{table*}[!t]
\centering
\caption{Field-evidence catalog: published observations of a measured response of large computational load. Type: F = single fault event; S = aggregated series of events; R = long operating record or oscillation episode; D = controlled demonstration. Boundary: POI = grid side at or beyond the point of interconnection; MSB/R = inside the facility. CM = cryptocurrency mining; DC = data center. CILR = customer-initiated load reduction.}
\label{tab:events}
\footnotesize
\begin{tabularx}{\textwidth}{@{}l l l l >{\raggedright\arraybackslash}p{2.0cm} Y Y l@{}}
\toprule
ID & Type & Load & Boundary & Date, system & Trigger & Reported load response & Source \\
\midrule
E1 & F & DC & POI & 10 Jul 2024, Dominion & 230\,kV SLG; 3 recloses per terminal, 6 fault applications in $\sim$82\,s & $\sim$1{,}500\,MW sustained for hours; co-located sites differed & \cite{nerc_ir2025,nerc_wp_disturbance}\\
E2 & F & DC & POI & 17 Feb 2025, Dominion & 230\,kV SLG; 4 depressions; 5.9-cycle clearing & $\sim$1{,}800\,MW; 18 substations, 44 transformers; mostly after 3rd depression & \cite{nerc_wp_disturbance}\\
E3 & F & DC & POI & 3 May 2025, Dominion & L--L fault; 3 depressions in $\sim$1\,s & $\sim$540\,MW, reconnected in 15--20\,min & \cite{nerc_wp_disturbance}\\
E4 & F & DC & POI & 19 Jun 2025, Dominion & 3-phase fault evolving to high-$R$ SLG & $\sim$1{,}300\,MW, much reconnected $<$1\,min & \cite{nerc_wp_disturbance}\\
E5 & S & CM, other & POI & 2023--2026, ERCOT & 32 events $\geq$100\,MW (series; may include E6) & 28 crypto-mining; largest non-computational 818\,MW (VFDs) & \cite{nerc_wp_disturbance}\\
E6 & F & CM & POI & 24 Jun 2025, ERCOT & LLG fault, $<$0.7\,pu for $>$20\,ms & Two crypto loads $-$52/$-$38\,MW; current 150--160\,\%; recovery 5\,min vs 2\,h & \cite{nerc_wp_disturbance}\\
E7--E9 & F & DC & POI & Feb--Mar 2025, AEP (Ohio) & 138\,kV faults, failed recloses & 428 of 814\,MW; 25\,MW; 227\,MW; dip-count logic decisive & \cite{nerc_wp_disturbance}\\
E10--E13 & F & DC & POI & 2022--2025, EirGrid & 220\,kV SLG faults & 74--387\,MW (16--52\,\% of DC load); RoCoF up to 0.26\,Hz/s & \cite{nerc_wp_disturbance}\\
E14 & D & DC (UPS) & POI & 2018, Statnett & UPS fast-frequency-reserve pilot & Fast activation; 30\,s duration not sustained & \cite{statnett2018,alapera2019}\\
E15 & R & DC & MSB/R & 2017, Meta facility & Internal PSU/grid resonance & 49/71\,Hz at switchboard and rack; event count unknown & \cite{sun2022p1,sun2022p2}\\
E16 & R & DC & POI & $\sim$1\,yr record, Dominion & Post-outage ``power swings'' & $\sim$0.40\,Hz load duty cycling, not instability & \cite{mishra2026discerning}\\
E17 & R & DC & POI & $\sim$2\,h episode, Dominion & Once-per-second voltage sag as exogenous excitation & 14.7--14.8\,Hz limit cycle from a 10--11\,Hz mode; owner confirmed UPS input stage & \cite{s3_147hz}\\
E18 & F & DC & POI & 22 Jul 2026, Dominion (PJM) & 230\,kV line fault, correctly cleared & $\sim$3{,}800\,MW to backup, the largest in PJM; initial transfer raised voltages and triggered further transfers; ACE $+$3{,}928\,MW, 60.092\,Hz & \cite{wharton2026pjm}\\
E19 & F & DC & MSB & before Jan 2024, PNM (WECC) & Transmission fault cleared in 0.4\,s; manual reclose 85\,s later & $\sim$200\,MW OCP-design site rode through; switchboard voltage 0.50\,pu, current doubled (1{,}400$\rightarrow$2{,}800\,A); standard composite load model did not match & \cite{jenkins2024wecc}\\
\bottomrule
\end{tabularx}
\end{table*}

Most single events involve data centers in PJM and EirGrid, whereas the ERCOT series is dominated by cryptocurrency mining, whose protection logic need not transfer to data centers. Four observations from these records bear directly on model credibility.

\emph{Same site, opposite outcome.} One Dominion site shed about 20\,\% of its load at a minimum phase voltage of 0.7205\,pu for 67\,ms (E3), yet shed its full load at a milder 0.7538\,pu for 75\,ms (E4)~\cite{nerc_wp_disturbance}. The deeper, shorter dip produced the smaller response. A voltage--duration characteristic could in principle separate the two, so these two summary points do not by themselves exclude static thresholds; together with differences in voltage definition (minimum phase versus positive sequence) and possible setting changes between events, they show that voltage magnitude alone does not characterize the facility. The stronger evidence for protection memory is the dip counting below.

\emph{Dip counting.} In E2 and E9 most reductions followed the third voltage depression, and one AEP site that saw four sags below 0.9\,pu did not reduce at all because its logic counted dips differently~\cite{nerc_wp_disturbance}. Facilities have since changed settings (for example, counting only dips longer than six cycles and below 0.85\,pu), so a model parameterized before such a change is no longer the installed facility.

\emph{Sites interact.} In E18 the initial transfers raised network voltages, and the resulting overvoltage triggered further transfers~\cite{wharton2026pjm}. Aggregate behavior is therefore not the sum of independently modeled sites; a model that represents each facility's undervoltage logic but not its overvoltage response, or that is checked one site at a time, cannot reproduce the cascade.

\emph{Measurement changes the diagnosis.} Dominion's long-term synchrophasor analysis showed that apparent post-outage power swings were $\sim$0.40\,Hz workload duty cycling rather than a stability problem~\cite{mishra2026discerning}. The interpretation came from measurement, not from a model.

\subsection{Requirements now in force or proposed}
\begin{figure*}[!t]
\centering
\includegraphics[width=\textwidth]{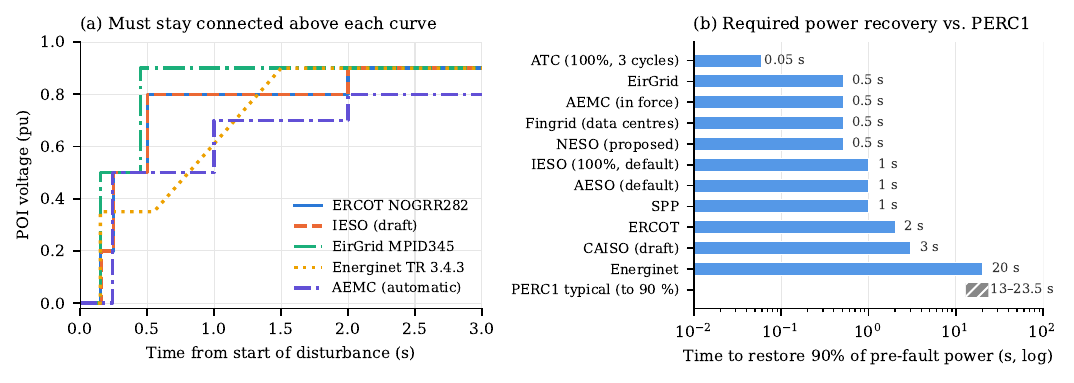}
\caption{(a) Undervoltage envelopes from five primary documents~\cite{ercot_nogrr282,ieso2026lcl,eirgrid2026mpid345,energinet2024tr343,aemc2026final} (the AEMC envelope is read from its published figure). (b) Time allowed to restore 90\,\% of pre-disturbance power after voltage recovers (100\,\% for ATC and IESO), compared with the time to 90\,\% under typical PERC1 settings, reconnection delay plus 90\,\% of a linear ramp (13--23.5\,s)~\cite{kosterev2026perc1}. Only the 90\,\% point is plotted; Energinet's additional 80\,\%-within-5\,s point is also missed by typical PERC1 settings.}
\label{fig:env}
\end{figure*}

Table~\ref{tab:reqs} compares fifteen requirement documents, whose status ranges from rules in force to drafts and position papers, from three U.S. markets, two U.S. transmission owners, two Canadian operators, six European operators, the European TSO association and the Australian market, and Fig.~\ref{fig:env} plots the envelopes and recovery times that can be compared directly. Australia's rule for inverter-based loads above 30\,MW took effect on 12 November 2026~\cite{aemc2026final}; Denmark's has been in force since November 2024~\cite{energinet2024tr343}; Texas's has applied since 1 August 2026, after approval by the state regulator~\cite{ercot_nogrr282,ercot2026puct}; Ireland's was approved by its regulator on 22 September 2026 and took effect five weeks later~\cite{eirgrid2026mpid345,cru2026mpid345}; most others are drafts or proposals, which we use to identify trends, not as existing obligations.

\begin{table*}[!t]
\centering
\caption{Large-load ride-through and recovery requirements (status as of November 2026). ``---'' = not specified in the documents reviewed. $^\dagger$Read from a published figure. $^\ddagger$From the ESIG international survey~\cite{esig2026perf}, not the primary document.}
\label{tab:reqs}
\scriptsize
\begin{tabularx}{\textwidth}{@{}>{\raggedright\arraybackslash}p{2.6cm} >{\raggedright\arraybackslash}p{2.0cm} Y >{\raggedright\arraybackslash}p{2.9cm} >{\raggedright\arraybackslash}p{3.1cm}@{}}
\toprule
Jurisdiction & Status & Voltage ride-through (stay connected) & Repeated disturbances & Active-power recovery\\
\midrule
ERCOT NOGRR\,282~\cite{ercot_nogrr282,ercot2026puct} & In force, Aug 2026 & $<$0.2\,pu 0.15\,s; 0.2--0.5\,pu 0.25\,s; 0.5--0.8\,pu 0.5\,s; 0.8--0.9\,pu 2\,s; 1.1--1.2\,pu 2\,s & --- & $\geq$90\,\% within 2\,s\\
CAISO~\cite{caiso_straw} & Straw proposal, Jun 2026 & $<$0.25\,pu 0.15\,s; $<$0.5\,pu 1.2\,s; $<$0.7\,pu 3\,s; $<$0.9\,pu 6\,s & Durations cumulative within any 10\,s & $\geq$90\,\% within 3\,s\\
SPP~\cite{spp_rt2026} & Proposed, 2026 & Keep consuming at 0.5--0.8\,pu, power reduced in proportion to voltage; UPS transfer permitted below 0.5\,pu & No disconnection triggered by multiple reclosing dips & 1\,s, extendable for weak grids\\
ATC$^\ddagger$ & Final & Envelope specified & 3 events within 10\,s & Full within 3 cycles, no transfer to backup\\
AESO$^\ddagger$ & Final, Jun 2026 & Envelope specified; current blocking not permitted & One or more faults within 10\,s & 1\,s default, up to 10\,s\\
IESO (Ontario)~\cite{ieso2026lcl} & Draft v1.0, May 2026 & $<$0.2\,pu 0.16\,s; 0.2--0.5\,pu 0.25\,s; 0.5--0.8\,pu 0.5\,s, power reduced only in proportion to voltage; no UPS transfer for overvoltage up to 1.05\,$V_{\max}$ & Two disturbances within 30\,s (failed reclose) & 100\,\% within 1\,s (default); no ramp limit; RoCoF 5\,Hz/s\\
Southern Company~\cite{soco2026tcll} & In use, v1.4 May 2026 & Facility submits settings for review; no fixed envelope & --- & --- (oscillation limits instead)\\
EirGrid~\cite{eirgrid2026mpid345,soni2026,cru2026mpid345} & Approved by regulator, Sep 2026 & 0\,pu to 0.15\,s; 0.5\,pu to 0.45\,s; 0.9\,pu thereafter; RoCoF 1\,Hz/s & --- & $\geq$90\,\% within 0.5\,s of voltage $\geq$0.9\,pu\\
Energinet TR 3.4.3~\cite{energinet2024tr343} & In force, Nov 2024 & 0\,pu to 0.15\,s; 0.35\,pu to 0.55\,s; linear to 0.9\,pu at 1.5\,s; current blocking allowed & --- & 80\,\% in 5\,s; 90\,\% in 20\,s; 95\,\% in 30\,s\\
Fingrid$^\ddagger$ & 2026 specification & Envelope specified; current blocking below 0.5\,pu & 10 bolted faults of 100\,ms within 90\,s & Data centers 90\,\% within 0.5\,s\\
TenneT$^\ddagger$ & Specified & Envelope specified; current blocking below 0.5\,pu & 3 LVRT and 5 HVRT events within 30\,min & Pre-fault current restored after clearance\\
RTE (France)$^\ddagger$ & In force, 2024 & Envelope specified & --- & ---\\
NESO (Great Britain)~\cite{neso2025gcdf} & Proposal, Nov 2025 & Mirrors generator fault ride-through & --- & $\geq$90\,\% within 0.5\,s of voltage $\geq$0.9\,pu\\
ENTSO-E~\cite{entsoe2025cnc} & Position paper, Dec 2025 & 0\,pu to 0.15\,s, then linear to $U_{rec2}$ at 3\,s; overvoltage 1.3\,pu 0.1\,s, 1.2\,pu to 60\,s & --- & 80--100\,\% within 0.5--5\,s, set by each TSO\\
AEMC (Australia)~\cite{aemc2026final} & In force, 12 Nov 2026 & May cease below 0.5\,pu for up to $\sim$0.24\,s; $\geq$0.5\,pu to 1\,s; $\geq$0.7\,pu to 2\,s; $\geq$0.8\,pu to 10\,s$^\dagger$; power may not fall more than in proportion to voltage & --- & $\geq$90\,\% within 0.5\,s, for disturbances up to 240\,ms\\
\bottomrule
\end{tabularx}
\end{table*}

Three features of these rules matter for modeling. First, many of them carry \emph{memory}. CAISO makes ride-through durations cumulative over any 10\,s window~\cite{caiso_straw}; SPP prohibits disconnection caused by multiple reclosing dips~\cite{spp_rt2026}; IESO requires two disturbances within 30\,s to be ridden through~\cite{ieso2026lcl}; ATC, AESO, Fingrid and TenneT specify multiple events within windows from 10\,s to 30\,min~\cite{esig2026perf}. A model that cannot remember the preceding dip, which is the dip-counting behavior observed in the field, cannot be used to show compliance with any of them. Second, recovery deadlines cluster at 0.5--3\,s (Fig.~\ref{fig:env}b), whereas the reconnection delay of about 4--10\,s plus a ramp of about 10--15\,s used as typical PERC1 settings~\cite{kosterev2026perc1}, which reach 90\,\% after 13--23.5\,s, meets none of them: Denmark's rule allows 20\,s for 90\,\% but also requires 80\,\% within 5\,s, which even the fastest typical settings (80\,\% at about 12\,s) miss. Third, several rules constrain \emph{how} power may fall during the sag, in proportion to voltage rather than to zero~\cite{spp_rt2026,ieso2026lcl,aemc2026final}, which a model with a single dropout threshold cannot express.

In North America these requirements are becoming standards: under NERC's Project 2026-02, the Board of Trustees adopted the first reliability standards for computational loads on 26 October 2026, for filing with FERC~\cite{appa2026nerc}, including one on interconnection, studies and modeling data that draws on the generator model-verification standard MOD-026~\cite{meier2026nercclo}; they are not yet enforceable. Practitioner accounts describe the same drivers and the same reliance on studies to set the values~\cite{conto2025,meier2026,rashwan2026,thomas2026}, propose storage at the facility as the route to compliance~\cite{hussein2026}, and note that most ride-through analyses so far rest on ad hoc simulations~\cite{norouzi2026dualpq}. The rules share a three-part structure: stay connected within an envelope, hold an active-power floor during the sag, and restore consumption within a deadline~\cite{wang2026solvrt}. Whether a facility meets them depends on device-level protection settings that planners rarely see, and no accepted equipment-level test framework yet maps grid requirements to immunity standards such as ITIC, IEEE~1668 or SEMI~F47~\cite{epri2026rt}. In Europe the gap is also legal: the Demand Connection Code does not cover data centers below the transmission level, its proposed amendment adds no ride-through clauses for them~\cite{acer2025dc}, and its adoption has been deprioritized, so ENTSO-E urges TSOs to act nationally~\cite{entsoe2025cnc,xie2026eenergy}. The rules also disagree: recovery deadlines range from three cycles (full recovery, as reported in a secondary survey) to 20\,s (90\,\%), more than two orders of magnitude (Fig.~\ref{fig:env}b), and one source still lists ERCOT's deadline as 1\,s from an earlier draft~\cite{esig2026perf}, whereas the rule in force says 2\,s~\cite{ercot_nogrr282}. A model shown to reproduce compliance with one envelope is not shown to reproduce compliance with another.

\subsection{How compliance is demonstrated}
Table~\ref{tab:compliance} asks a different question of the documents in Table~\ref{tab:reqs} that describe, at least in outline, how a submitted model or the installed facility is checked: what evidence does each operator accept that the submitted model behaves like the installed facility? Within the procedures these documents establish, the answer is almost always another model. Model-to-model benchmarks, equipment type tests and post-commissioning monitoring serve different purposes, and only the last observes the installed facility. EirGrid assesses compliance through studies, with facilities submitting self-assessments that ``may be supported by'' their own equipment tests, and its April 2026 framework proposal states that ``no grid level electrical testing will be undertaken''~\cite{eirgrid2026framework}; the regulator endorsed the framework with amendments~\cite{cru2026mpid345}. It also computes, from ``model data provided by Data Centres'' and its own simulations, a demand-loss factor that caps the aggregate data-center demand the Irish system will accept~\cite{eirgrid2026framework}. A national operating limit therefore rests on facility models that the framework does not require to be compared with a measured facility response; the documents do not show whether any such comparison has been made privately. IESO requires RMS models for two platforms, a PSCAD model, and a report benchmarking the RMS models against the EMT model, which is a model-to-model check, together with type or commissioning tests showing that the installed equipment meets the requirements~\cite{ieso2026lcl}. Even for inverter-based generators, where such benchmarking is routine, the agreement obtained depends on how plant-controller limits are mapped between platforms and on the operating point~\cite{cheerla2026emtrms}. Southern Company comes closest to validation: before declaring full readiness it runs a performance-validation period in which monitoring and simulations check ride-through performance and ``load modeling data accuracy'' ``to the extent possible'', without a stated acceptance criterion~\cite{soco2026tcll}. NERC's final technical reference describes the same practice in general terms: during commissioning, monitored field data ``may be compared directly to the modeled response'', and any deviation is corrected by updating model parameters, control settings or mitigation~\cite{nerc_trd}. By the definitions of Section~\ref{sec:framework}, that is calibration to the commissioning record; it becomes validation only if the corrected model is then frozen and tested against a later, unseen disturbance. Australia has asked its operator to write modeling guidelines for loads; one industry submission to that process argued that validating data-center models to the standard of generator models ``may be infeasible'', given the difficulty of benchmarking against measured responses~\cite{aemc2026final}.

A developer's account to the ESIG workshop puts the practical difficulty plainly: no field data exist before energization, bench tests cover converters in isolation rather than the whole site, few events are available for benchmarking, and validation ownership between a shell developer and its tenant is unclear; verification and validation are ``well established for generators \ldots\ but still largely undefined for large loads''~\cite{syed2026esig}. The same transmission owner that monitors large-load model accuracy only ``to the extent possible'' requires inverter-based generators to submit unit models validated against type or hardware-in-the-loop tests for large-signal disturbances, followed by plant-level validation before commercial operation~\cite{soco2025ibrmodel}. The generator standard itself is less prescriptive than the framework proposed here: MOD-026-2, approved by FERC on 19 February 2026, requires documentation comparing the model with ``the measured behavior during a staged test or system disturbance'' (Requirement R2, Part 2.3; Requirement R3, Part 3.3 for EMT models), and leaves acceptability to the transmission planner, judged against model requirements the planner sets (Requirements R1 and R5); it does not require the compared event to be withheld from parameter tuning, nor does it fix a numerical tolerance~\cite{nerc_mod026}. Gates G5 and G7 (Section~\ref{sec:framework}) therefore go beyond current generator practice as well; we propose them as recommendations, not as an existing requirement. The CIGRE large-load task force recommends the same for loads: model validation ``should be undertaken routinely, similar to generator model validation, using hardware-in-the-loop testing and/or actual events with high-resolution data''~\cite{cigre2026cse41}.

\begin{table*}[!t]
\centering
\caption{What evidence each requirement accepts that a submitted model represents the installed facility.}
\label{tab:compliance}
\footnotesize
\begin{tabularx}{\textwidth}{@{}>{\raggedright\arraybackslash}p{2.6cm} Y Y Y@{}}
\toprule
Requirement & Model required & Evidence that the model matches the facility & Field verification\\
\midrule
NERC Level 3 alert~\cite{nerc_l3alert} & PERC1 or equivalent as a minimum & Parameters from OEM outreach and an operator questionnaire~\cite{nerc_questionnaire} & None specified\\
EirGrid~\cite{eirgrid2026framework} & Studies per study assessment guides; model data feed a system-wide demand cap & Self-assessment, ``may be supported by'' own equipment tests & ``No grid level electrical testing''\\
IESO~\cite{ieso2026lcl} & PSS/E and DSA composite or user models; PSCAD model & RMS models benchmarked against the EMT model (model-to-model) & Type tests or on-site commissioning tests of equipment\\
Southern Company~\cite{soco2026tcll} & As-planned, as-built and as-left models; simulations before energization & Performance-validation period checks ``load modeling data accuracy'' ``to the extent possible''; no acceptance criterion & Post-commissioning monitoring; re-validation after material change\\
AEMC~\cite{aemc2026final} & Modeling guidelines for loads to be developed by AEMO & Under development & Operator may request tests\\
Energinet~\cite{energinet2024tr343} & Compliance partly by simulation with the TSO & --- & Compliance tests may be requested\\
\emph{Contrast: IBR generators}~\cite{soco2025ibrmodel} & Unit and plant models & Unit models validated against type or HIL tests for large-signal disturbances & Plant-level validation before commercial operation\\
\bottomrule
\end{tabularx}
\end{table*}

\section{Model Landscape}\label{sec:models}
\begin{figure*}[!t]
\centering
\includegraphics[width=\textwidth]{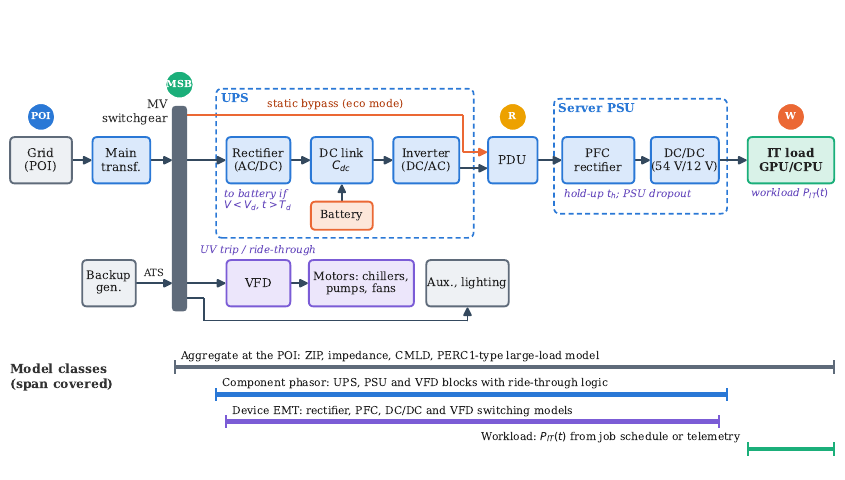}
\caption{Reference electrical model of a data center: the IT power path through the UPS and the server power supply, the cooling and auxiliary loads, the control and protection logic (purple), the measurement boundaries (POI, MSB, R, W, as in Fig.~\ref{fig:path}), and the part of the facility that each model class represents.}
\label{fig:model}
\end{figure*}

\subsection{A reference structure for data-center dynamic models}\label{sec:refmodel}
Fig.~\ref{fig:model} places the model classes of this review on one power path. This subsection writes the common elements as equations so that the evidence for each can be stated against a named quantity. The forms are standard; none is proposed here as a new model.

\emph{Static aggregate.} A ZIP load writes active power at the POI as
\begin{equation}
P(V)=P_0\left[p_z\left(\frac{V}{V_0}\right)^{2}+p_i\,\frac{V}{V_0}+p_p\right],
\label{eq:zip}
\end{equation}
where $p_z+p_i+p_p=1$, with a matching expression for $Q$~\cite{milanovic2013,arif2018}. A power-electronic load regulates its output, so near nominal voltage $p_p\approx1$ and the input current rises as voltage falls, $I\approx P_0/V$, until a current limit or a dropout threshold is reached~\cite{sun2019pfc}.

\emph{Energy buffer.} A dip is ridden through from the DC-link capacitor of the server power supply or the UPS, for at most
\begin{equation}
t_h=\frac{C_{dc}\left(V_{dc,0}^{2}-V_{dc,\min}^{2}\right)}{2P_{dc}},
\label{eq:holdup}
\end{equation}
where $V_{dc,\min}$ is the lowest DC voltage at which the downstream converter still regulates; the expression assumes constant-power discharge and no input current during the sag. Battery-backed stages replace~\eqref{eq:holdup} by the battery's stored energy and discharge limit.

\emph{Transfer and dropout logic.} The grid-side connection state $u\in\{0,1\}$ of a UPS or PSU group changes when
\begin{equation}
u(t)=0\ \ \text{if}\ \ V(\tau)<V_d\ \ \forall\,\tau\in[t-T_d,\,t],
\label{eq:dropout}
\end{equation}
with pickup voltage $V_d$ and delay $T_d$. The group returns ($u=1$) after $V>V_r$ for $T_r$ and reloads at ramp rate $R$:
\begin{equation}
P(t)=\min\!\left\{P_0,\;R\,(t-t_r)\right\},\quad t\ge t_r,
\label{eq:ramp}
\end{equation}
which is the structure of the cease, reconnect and ramp parameters in the PERC1-type large-load model~\cite{kosterev2026perc1,odunlami2026perc1}. Equation~\eqref{eq:ramp} reloads a group from zero; with partial cessation it applies to the ceased share of each group.

\emph{Composite aggregate.} In the composite load model the power-electronic fraction is disconnected progressively between two voltages,
\begin{equation}
f_e(V)=\begin{cases}1, & V\ge V_{d1}\\[2pt] \dfrac{V-V_{d2}}{V_{d1}-V_{d2}}, & V_{d2}<V<V_{d1}\\[4pt] 0, & V\le V_{d2}\end{cases}
\label{eq:cmld}
\end{equation}
so that~\eqref{eq:dropout} applied to many units with spread thresholds is approximated by a linear characteristic~\cite{wecc_cmld_spec,nerc2016dlm,kosterev2008,sleiman2025cmld}.

\emph{Facility at the POI.} The power seen by the grid sums the IT path, the cooling motors and the auxiliaries,
\begin{equation}
P_{\mathrm{POI}}=\sum_{k}u_k\,\frac{P_{IT,k}}{\eta_k}+P_m(V,\omega)+P_{\mathrm{aux}}(V)+P_{\ell},
\label{eq:poi}
\end{equation}
where $k$ indexes UPS or PSU groups with efficiency $\eta_k$, $P_\ell$ is the network loss, $P_m$ is the motor (cooling) load through its drives, and the IT demand follows the workload,
\begin{equation}
P_{IT}(t)=N\left[P_{\mathrm{idle}}+\left(P_{\mathrm{pk}}-P_{\mathrm{idle}}\right)\rho(t)\right],
\label{eq:workload}
\end{equation}
with $N$ accelerators, idle and peak power per accelerator, and utilization $0\le\rho(t)\le1$. Each evidence cell in Section~\ref{sec:matrix} concerns one term of~\eqref{eq:zip}--\eqref{eq:workload}: a ride-through claim tests $V_d$, $T_d$ and $t_h$; a reconnection claim tests $V_r$, $T_r$ and $R$; an oscillation claim tests $\rho(t)$ and its spectrum.

\subsection{Model classes and their validation evidence}
Table~\ref{tab:models} summarizes the model classes in current use and the strongest validation evidence found for each in this review. Four observations stand out.

\emph{Standard library models.} The electronic component of the composite load model (CMLD) represents data-center load as constant power with one or two undervoltage dropout thresholds and an immediately restoring fraction; it has no time delays for dropout or reconnection, and NERC's technical reference states that using it ``will result in optimistic results that will conceal valid reliability impacts''~\cite{nerc_trd}. Even so, CMLD remains the submission format in some footprints, parameterized site by site~\cite{entergy2026,spp_ops2026}. The older IND\_SRF representation is described as ``no longer adequate'' for data-center response to voltage sags~\cite{kosterev2026perc1}. The first of two published comparisons we found between a standard library model and a measured data-center event points the same way: for a $\sim$200\,MW Open Compute facility that rode through a 0.50\,pu sag while its switchboard current doubled, the typical PSLF data-center composite load model ``does not match the observed behavior of the real datacenter''~\cite{jenkins2024wecc}. A second comparison, by a university team with utility co-authors, used single-phase waveforms of real voltage-sag events at data centers: with the default parameters of the composite load model, the simulated current differed from the record by 112\,\% on a deep sag (0.55\,pu for 0.22\,s) and by 22\,\% on a shallow one~\cite{sleiman2025cmld}. Machine-learning inversion over 2{,}806 sampled parameter sets, with the fault duration and impedance also tuned to the record, reduced the error to 9\,\% and 6.6\,\%. This is calibration: each event received its own fitted set, in a generic PSS/E example network rather than the facility's own, and the ``validation on other events'' repeats the fit rather than testing a frozen set on an unseen event~\cite{sleiman2025cmld}. Both comparisons agree that default library parameters do not reproduce measured data-center sag responses. NERC's data-collection questionnaire was launched in response to such disturbances that ``have not followed existing model responses''~\cite{jenkins2024wecc,nerc_questionnaire}.

\emph{PERC1.} The Power Electronic Reconnecting and Ceasing model adds ceasing and reconnection thresholds, timers and a reconnection ramp, and the NERC Level 3 alert directs planners to use it, or a model with equivalent or better capabilities, as a minimum~\cite{nerc_l3alert}. Its parameters are currently obtained from OEM outreach and a non-intrusive operator questionnaire~\cite{kosterev2026perc1,nerc_trd,nerc_questionnaire}. We found no published study that fits PERC1 to a measured data-center event. The only identification study uses synthetic PMU records and proves that one disturbance cannot identify every threshold--delay pair~\cite{odunlami2026perc1}. A successor model is being developed to represent emerging ride-through requirements~\cite{kosterev2026perc1}. A related model, LDEV1, added to one commercial tool in 2025 and derived from EPRI's aggregated EV-charger model, divides the load into paths with independent cease-and-reconnect logic and adds washout dynamics and current limits~\cite{powerworld_ldev1,javadi2026review}; we found no data-center study that uses or validates it, so it has no row in the matrix.

\emph{EMT models.} Two public EMT libraries exist. The ERCOT/TAMU PSCAD model is explicitly ``not validated against actual hardware''~\cite{ercot_pscad}; the PNNL library benchmarks its average-value models against switching models and its EMT data-center model against PERC1~\cite{pnnl_emt}, which are model-to-model checks. At device level, an EMT model of a cryptocurrency-mining power supply reproduced the trip/no-trip map of a laboratory miner at a few sag levels~\cite{samanta2024emt}, the laboratory validation that NERC's technical reference also cites~\cite{nerc_trd}. Industry practice is similar: PSU models are tuned against hardware-in-the-loop tests of physical supplies, with control and protection settings adjusted until the responses match~\cite{zamani2026esig}, which is calibration of a component rather than validation of a facility. Machine-learning surrogates inherit the same limitation: a GRU surrogate for UPS ride-through screening is referenced only to the EMT model it emulates, and, according to its abstract, a favorable single-seed result did not hold across seeds~\cite{s1_surrogate}. The only grid-event comparison we found for an EMT facility model replays a single ERCOT partial-tripping event of a mining facility using fault-recorder data, with the tripped fraction set to match the record~\cite{yazdanpanah2026partial}.

\emph{Detailed and workload-driven models.} Component-level phasor models~\cite{jimenezruiz2026,gyang2026tpwrs,arifin2026,ross2026,lu2026} and workload-calibrated models~\cite{lu2026pattern,sreejith2026,silva2026} increasingly draw on real data, but that data is usually IT telemetry (Fig.~\ref{fig:path}, boundary W) or component tests, not POI disturbance records. Data-driven models of cryptocurrency mining have been trained on EMTP-generated fault scenarios~\cite{samanta2024} or on device-level identification~\cite{ahmad2025}, so their reference is again a model or a device. Converter-based emulators of a data-center power system have been built and tested in a laboratory grid emulator~\cite{sun2021,sun2022}, which is the strongest hardware evidence for facility-level dynamics but still not a commissioned site. Utility practice is moving from generic to configured models: PJM has, for the first time, represented data-center loads in parts of its stability assessment, with ride-through capability configured per study~\cite{tang2026}, and EPRI's comparison of aggregate, detailed, phasor and EMT load representations shows how much the conclusion depends on that choice~\cite{ramasubramanian2025}; the same holds in steady-state distribution studies, where the mix of constant-power, constant-current and constant-impedance load changes voltage and loss results~\cite{bindal2024npsc}. Static extensions such as ZIP-E~\cite{reyes2024}, which adds electronic-load terms without protection logic, and a steady-state equivalent-circuit model for power-flow studies~\cite{ali2026steady} are evaluated only in simulation. System-impact studies of voltage stability, small modular reactor supply, regional dynamics and adversarial load manipulation~\cite{miah2026,badakhshan2026,hossain2025,zhabelova2015,barati2025} necessarily adopt one of these model classes and inherit its evidence; one multi-timescale study is unusual in publishing a fidelity table that prevents its results from being read beyond what its models support~\cite{shamseldein2026}. Measured data, where used, are mostly steady-state or IT-side: a telecom data center's harmonic load model was compared with on-site power-quality measurements~\cite{zubi2024}, and ``real-world'' datasets in frequency studies are scheduler and GPU telemetry~\cite{aghadinuno2026}.

\emph{A historical contrast.} The composite load model was itself developed by fitting and then checking model structures against recorded WECC events such as the 1996 outages and the 2000 oscillation, with field monitoring installed for the purpose~\cite{kosterev2008,nerc2016dlm}. Its electronic component has never been through the equivalent process for data centers. Load modeling in general has long lagged: an international survey found that about 70\,\% of utilities used only static load models and that parameters had been updated within five years in only about 40\,\% of cases~\cite{milanovic2013}, and reviews describe common load models as unable to represent emerging loads~\cite{arif2018,harzevili2026}. Data centers inherit this gap at a much larger scale per connection.

\begin{table*}[!t]
\centering
\caption{Data-center load model classes and the strongest validation evidence found. Tiers: T1 grid-side record; T2 facility-internal measurement; T2w IT-workload telemetry; T3 laboratory/device; T4 OEM or questionnaire parameters; T5 simulation or model-to-model only.}
\label{tab:models}
\footnotesize
\begin{tabularx}{\textwidth}{@{}l l Y Y l@{}}
\toprule
Model class & Domain & What it represents & Strongest validation evidence found & Tier \\
\midrule
ZIP / constant power & Phasor & Static $P(V)$, $Q(V)$ & Event records show near-constant $P$ during ride-through (qualitative)~\cite{nerc_trd} & T1 (qual.)\\
CMLD electronic component & Phasor & Constant $P$; $V_{d1}$, $V_{d2}$ dropout; immediate partial restart; no delays & Judged optimistic by NERC~\cite{nerc_trd}; site-specific parameterization in use~\cite{entergy2026}; typical PSLF data-center model reported not to match a measured sag response~\cite{jenkins2024wecc}; default parameters 22--112\,\% waveform error, per-event refit 6.6--9\,\%~\cite{sleiman2025cmld} & T2 (mismatch; calib.)\\
IND\_SRF & Phasor & Industrial load representation & Declared inadequate for sag response~\cite{kosterev2026perc1} & ---\\
PERC1 & Phasor & Cease/reconnect thresholds, timers, ramp & OEM/questionnaire parameters~\cite{kosterev2026perc1}; identifiability shown on synthetic PMU data only~\cite{odunlami2026perc1} & T4\\
LDEV1 & Phasor & Washout dynamics, current limits, three ceasing paths with separate cease/reconnect logic & None found for data centers; derived from an EV-charger model~\cite{powerworld_ldev1} & ---\\
Component-level phasor models & Phasor & IT, UPS, cooling, backup transitions & Component (server thermal/power) checks against experiments~\cite{gyang2026tpwrs}; system response simulated~\cite{jimenezruiz2026} & T3/T5\\
Workload-calibrated models & Phasor & Stochastic or pattern-matched demand & Calibrated to IT-utilization traces~\cite{lu2026pattern} & T2w\\
Impedance (PSU/PFC) models & Small-signal & Input impedance, internal resonance & Laboratory PSU tests and switchboard measurements~\cite{sun2022p1,sun2022p2} & T2/T3\\
EMT facility models & EMT & Converters, UPS, protection, internal network & Not hardware-validated~\cite{ercot_pscad}; model-to-model benchmarks~\cite{pnnl_emt}; single-event replay with imposed trip fraction~\cite{yazdanpanah2026partial} & T5 / T1 (calib.)\\
EMT device models & EMT & Single PSU/miner & Trip map vs.\ laboratory miner at a few sag levels~\cite{samanta2024emt}; PSU settings tuned to HIL tests~\cite{zamani2026esig} & T3 (calib.)\\
\bottomrule
\end{tabularx}
\end{table*}

\subsection{Data-driven and AI-based models}\label{sec:datadriven}
Data-driven methods are often proposed as the way around the parameter problem: if a model's structure or parameters can be learned from measurements, the gap between library models and measured behavior should close. Table~\ref{tab:datadriven} codes the 70 studies of data-driven and AI-based load models found by the update search (Section~\ref{sec:method}) by family and by the reference object of their strongest test. Four findings follow.

\begin{table}[!t]
\centering
\caption{Data-driven and AI-based load-model studies from the update search, by the reference object of their strongest test (tiers as in Table~\ref{tab:models}). Held-out: studies that test on events withheld from fitting, measured\,/\,simulated.}
\label{tab:datadriven}
\footnotesize
\setlength{\tabcolsep}{2pt}
\begin{tabular}{@{}l c c c c c c@{}}
\toprule
Family & $n$ & T1 & T2/T2w & T3 & T5 & Held-out\\
\midrule
ML identification of standard structures & 30 & 9 & 0 & 0 & 21 & 0\,/\,1\\
Learned or free-form structures & 16 & 1 & 1 & 1 & 13 & 0\,/\,6\\
Converter and device surrogates & 10 & 1$^{a}$ & 0 & 4 & 5 & 3\,/\,0\\
Data-center and AI-workload models & 14 & 0 & 8 & 0 & 6 & 0\,/\,1\\
\midrule
Total & 70 & 11 & 9 & 5 & 45 & 3\,/\,8\\
\bottomrule
\multicolumn{7}{@{}p{0.95\columnwidth}@{}}{\footnotesize $^{a}$Railway traction converter. No T1 record in the table is of a data center.}
\end{tabular}
\end{table}

\emph{The methods are maturing faster than the evidence.} Machine-learning identification of standard structures, from deep reinforcement learning for the WECC composite load model~\cite{wang2020ddqn,bu2020edrl} to Bayesian and diffusion-model estimators~\cite{qiu2026bayes,zhu2026diffusion}, and learned or free-form structures such as neural ODE and DAE modules~\cite{xiao2023neuralode}, neural dynamic equivalents~\cite{shen2024neudye}, symbolic regression~\cite{lin2026rlsr} and physics-informed networks~\cite{dorji2026pdkan}, are now common. Of the 46 studies in these two families, 34 are tested only in simulation. Ten use field records from utility substations, fault recorders and feeders~\cite{qi2022pmu,mitrentsis2021adn,paidi2020ocl,leinakse2021dfr,lin2024taiwan,avila2020cmld,bu2020edrl}; all are records of conventional aggregate loads, and none is of a data center.

\emph{Held-out testing has arrived, but on simulated events.} Eight studies test a trained model on events withheld from fitting: unseen fault locations~\cite{dorji2026pdkan,shen2024neudye}, unseen deep sags for a branch-resolved data-center model~\cite{lin2026identifiability} and unseen faults after a fitted composite load model is reinserted into a network~\cite{woo2026reinsertion}, among others. In all eight the withheld events are simulated. The three held-out tests on measured data are at device level: a DC--DC converter and an inverter-driven motor load in the laboratory~\cite{zhang2024lstmdcdc,varshney2025sysid}, and a railway traction converter. The procedure that gate G5 asks for is therefore in routine use; what is missing is a measured data-center event to withhold.

\emph{Learning sharpens the gate questions rather than removing them.} Several studies make the gates' concerns explicit. On identifiability (G4), different parameter sets can reproduce the same response~\cite{zhu2026diffusion,lu2025uniqueness}; information geometry shows which composite-load parameters measurement data can determine~\cite{transtrum2025tailoring}, and a Bayesian estimator quantifies the uncertainty that remains~\cite{qiu2026bayes}. On object identity (G3), a composite load model fitted by voltage replay, in which the measured voltage is imposed on an isolated load, can match the measured $P$ and $Q$ yet respond differently once reinserted into a network~\cite{woo2026reinsertion}. On acceptance criteria (G7), the response error between simulation and PMU data does not always indicate whether the load model is accurate~\cite{cicilio2020metrics}.

\emph{Data-center-specific work is recent and rests on telemetry or simulation.} Of the 14 data-center and AI-workload studies, seven use IT or meter telemetry, one uses measured electrical transients, including an open sub-second dataset of AI-training node power~\cite{elsayed2026dataset} and power signatures of HPC workloads~\cite{govind2023hpc}, and six are simulation-only, including a workload-aware extension of the composite load model~\cite{saroare2026dcclm}, a switched-DAE stability analysis~\cite{shenoy2026switched} and data-driven control against sub-synchronous resonance~\cite{ruan2025ssr}. The most complete is a branch-resolved model that fits its workload-to-power law to facility telemetry and calibrates cooling composition, UPS voltage gate and reconnect ramp under an explicit identifiability criterion; its sags, however, are simulated, and its authors note that field calibration would require branch-level measurements~\cite{lin2026identifiability}. The one study built on measured electrical transients at a data center reduces eight uncontrolled events to low-order representations, without stating the measurement boundary and, by its authors' account, without testing generalization to unseen events~\cite{acero2026processes}. At device level, neural surrogates inside EMT simulators~\cite{ventura2026pinn} and system identification from power hardware-in-the-loop experiments~\cite{guruwacharya2024phil} follow the pattern surveyed in~\cite{subedi2025overview}, which calls for a shared data repository to benchmark such models.

Coded by the rules of Appendix~A, none of these studies raises a data-center cell of Fig.~\ref{fig:matrix} above its recorded tier. The extension of~\cite{saroare2026dcclm}, which adds UPS state logic and workload profiles absent from the CMLD electronic component, is coded as a component-level phasor model, whose cells it matches; and~\cite{acero2026processes} represents measured events without modeling the facility's response to a class of grid disturbance. Data-driven models therefore need what physics-based ones need: measured data-center disturbance records, some withheld from fitting.

\section{A Context-of-Use Credibility Framework}\label{sec:framework}
\emph{Definitions.} We use five terms strictly. A comparison is \emph{independent} when no quantity of the model or of the simulated disturbance was adjusted to the event being compared. A model is \emph{frozen} when its structure, parameters and settings are fixed and recorded before that comparison. An event is \emph{held out} when it was not used to set any model quantity. A model is \emph{facility-specific} when its parameters describe one identified facility rather than a generic class. A \emph{successful validation} is a held-out, independent comparison of a frozen model that meets an acceptance criterion fixed in advance, for a stated context of use. Fitting parameters, or the disturbance itself, to an event is \emph{calibration}, however good the resulting agreement. Under these definitions the two default-parameter comparisons of the composite load model with measured events~\cite{jenkins2024wecc,sleiman2025cmld} are independent comparisons of a frozen generic model; both reported a mismatch, though neither stated an acceptance criterion; the per-event refits of~\cite{sleiman2025cmld} are calibration.

A model is not credible in general; it is credible for a stated use. Following the context-of-use logic of established verification and validation practice~\cite{asme_vv40}, we define the context of a data-center model claim as $C=(s,b,m,e,h,y,d)$: study purpose $s$, measurement boundary $b$ (Fig.~\ref{fig:path}), UPS/converter/control mode $m$, disturbance class $e$, time horizon $h$, endpoint $y$ (for example POI active power 3\,s after voltage recovery), and decision $d$ (for example compliance with a recovery requirement). Evidence for one context does not automatically transfer to another. The gates below adapt, rather than replace, established credibility practice: the context of use and the link between model risk and required evidence follow ASME V\&V~40~\cite{asme_vv40}, and the separation of verification, validation against referents, and stated limits of applicability follows NASA-STD-7009A~\cite{nasa7009}. What is specific to data centers is the content of each gate: the measurement boundaries of Fig.~\ref{fig:path}, the switching of UPS operating modes, protection logic with memory, and the identifiability limits of single events.

\begin{figure*}[!t]
\centering
\includegraphics[width=0.82\textwidth]{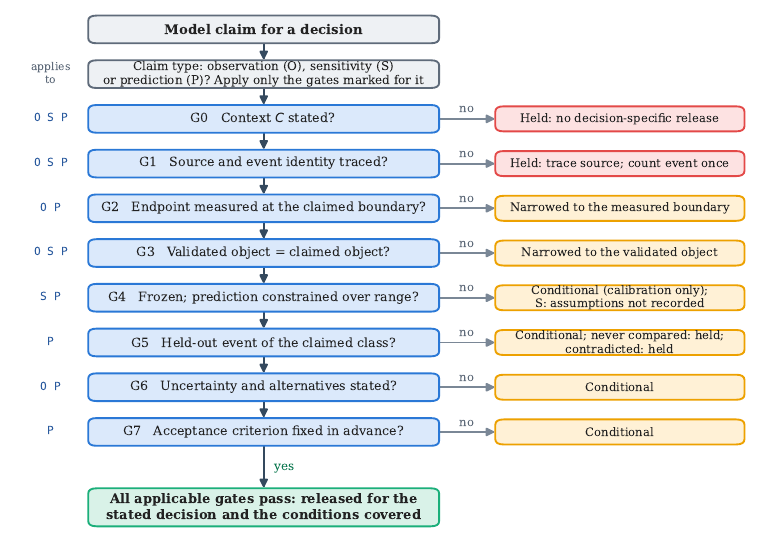}
\caption{The eight gates applied to a model claim. The claim type is identified first, and only the gates marked for it are applied (left column: O observation, S sensitivity, P prediction). A failed gate does not reject the study; it narrows the claim to what the evidence supports or holds it until the missing evidence exists. A prediction never compared with a measurement, or contradicted by one, is held.}
\label{fig:gates}
\end{figure*}

We then ask eight questions of every consequential claim (Fig.~\ref{fig:gates}). Each question either allows the claim, restricts it, or holds it.

\begin{enumerate}
\item[G0] \emph{Context.} Are $s,b,m,e,h,y,d$ stated? If not, no decision-specific release.
\item[G1] \emph{Source and event identity.} Is the evidence traced to an original source, site and dated event, so that repeated publications of one event are counted once?
\item[G2] \emph{Measurement boundary.} Was the endpoint measured at the boundary the claim concerns (POI vs.\ switchboard vs.\ workload telemetry), with sufficient bandwidth?
\item[G3] \emph{Validated object.} Is the object that was compared with measurements the same object the claim is about? Validating a power-system model, a single power supply, or a server thermal model does not validate a facility model.
\item[G4] \emph{Model lifecycle and predictive identifiability.} Were parameters frozen before the comparison? Do the calibration events constrain the \emph{decision-relevant output} over the claimed conditions? Individual parameters need not be identifiable, structurally or practically; what is required is that every parameter set consistent with the calibration data predicts the endpoint within the acceptance criterion over the claimed range. For PERC1, a single disturbance leaves at least one threshold--delay direction unidentifiable~\cite{odunlami2026perc1}; a fit to one event is calibration, not validation.
\item[G5] \emph{Held-out comparison.} Was the frozen model compared with a disjoint event, of the disturbance class the claim concerns?
\item[G6] \emph{Uncertainty.} Are uncertainties in measurement, parameters and operating state stated, including alternative mechanisms such as a workload change instead of a backup transfer~\cite{albustami2026attribution}?
\item[G7] \emph{Decision rule.} Is the acceptance criterion stated before the comparison, together with the consequence of a wrong acceptance?
\end{enumerate}
The labels follow an explicit rule. A claim is \textbf{held} if G0 or G1 fails, if a prediction about facility behavior has never been compared with a measurement (records that were not compared with the model do not count), or if the only measured evidence contradicts it: essential evidence is absent or adverse. If G2 or G3 fails, the claim is \textbf{narrowed} to the boundary and object actually measured, and the gates are re-applied to the narrowed claim; agreement at a power-supply terminal cannot substitute for a missing POI measurement. If G0--G3 pass but any of G4--G7 fails, the claim is \textbf{conditional}: some evidence supports it under stated assumptions, and the unresolved gates are listed with it. A claim is \textbf{released} only when every gate that applies to it passes (see claim types below), and only for the conditions the evidence covers. Agreement on one held-out event is \emph{event-specific}; extending it to a range of conditions requires G4 to show that the prediction is constrained over that range (\emph{condition-specific}); a release for a decision covers no more than that range; and credibility for a facility in general would require held-out agreement across its operating conditions. Where several gates fail, the most restrictive label applies (held, then narrowed, then conditional). Expert judgment remains in deciding whether a boundary or object is ``the same'' (G2, G3) and whether uncertainty is adequately stated (G6); the worked cases record those judgments. The framework is a proposed evaluation procedure, not a validated acceptance standard. The labels describe the evidence, not the quality of the underlying work: a careful simulation study can be held for a compliance claim while being fully adequate for a sensitivity claim.

\emph{Claim types.} Not every gate applies to every claim. An \emph{observation} states that a phenomenon occurred, with its cause, at a stated boundary; no model is involved, so G4, G5 and G7 do not apply. A \emph{sensitivity} claim states how a result depends on assumed facility behavior without asserting that the model represents a particular facility; it is conditional on its stated assumptions by construction, so G2 and G5--G7 do not apply, while G4 requires those assumptions to be recorded. A \emph{prediction}, which includes every compliance claim, states what a facility will do, and all eight gates apply. A prediction that has been compared with an independent measurement and contradicted by it is held for the disturbance class compared, whatever the other gates show. Table~\ref{tab:gateout} records the outcome of each gate for the worked cases of Section~\ref{sec:cases}.

\section{Evidence Matrix}\label{sec:matrix}
\begin{figure*}[!t]
\centering
\includegraphics[width=\textwidth]{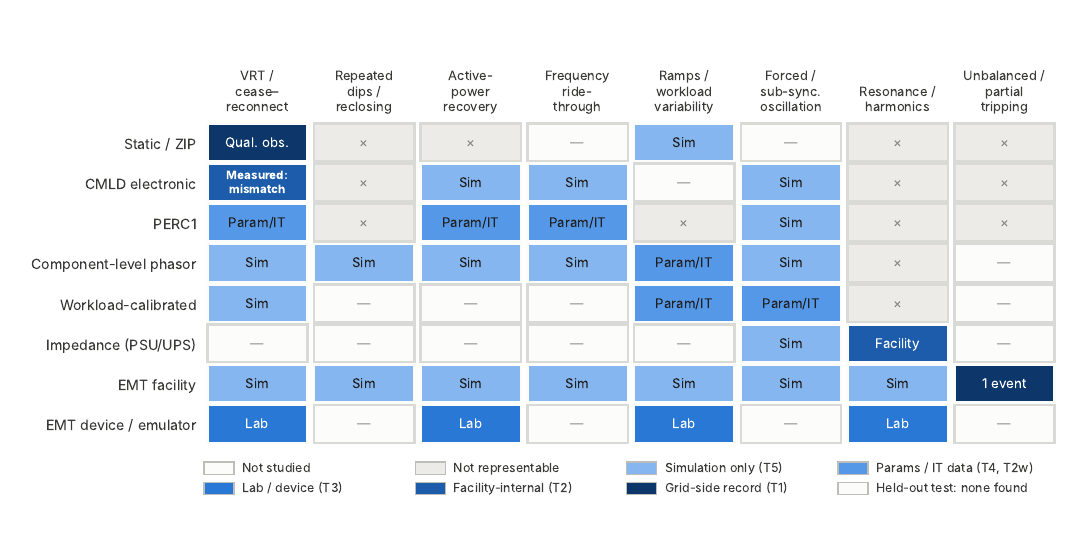}
\caption{Evidence matrix: strongest provenance of evidence, positive or negative, about whether each model class reproduces each phenomenon. ``Measured: mismatch'' marks the composite load model's mismatches with measured sag responses at default parameters (switchboard record~\cite{jenkins2024wecc}; boundary not stated~\cite{sleiman2025cmld}), coded conservatively at T2, and its per-event recalibration; ``1 event'' is a single-event replay of a mining facility with the tripped fraction set to match~\cite{yazdanpanah2026partial}; ``Qual. obs.'' is the qualitative observation of near-constant power during ride-through~\cite{nerc_trd}. No cell holds a successful held-out validation. Coding rules are in Appendix~A.}
\label{fig:matrix}
\end{figure*}

Fig.~\ref{fig:matrix} crosses eight of the ten model classes of Table~\ref{tab:models} (IND\_SRF and LDEV1 have no data-center study to code) with eight grid-relevant phenomena and records, for each cell, the strongest provenance of evidence found about whether a model of that class reproduces that phenomenon, whether that evidence supports the model or contradicts it. The unit is a model class and phenomenon, not a paper, so repeated publications from one program count once (gate G1). IND\_SRF, listed in Table~\ref{tab:models}, is omitted because it has been declared inadequate and we found no study that applies it to these phenomena. Cells are marked \emph{not representable} when the model structure cannot express the behavior at all (for example, a positive-sequence model and unbalanced partial tripping, or PERC1 and dip counting).

\begin{figure}[!t]
\centering
\includegraphics[width=\columnwidth]{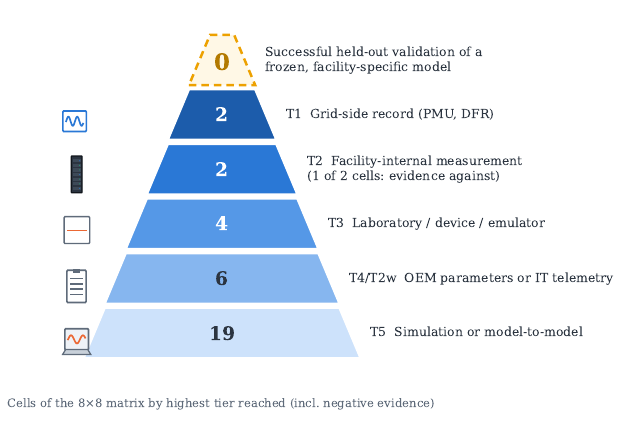}
\caption{The evidence pyramid of Fig.~\ref{fig:matrix}: number of model--phenomenon cells whose strongest evidence reaches each tier. Counts are by the highest provenance reached and include evidence against a model. The top level, a successful held-out validation of a frozen, facility-specific model, is a validation status rather than a provenance tier (Table~\ref{tab:provstatus}); it is empty.}
\label{fig:pyramid}
\end{figure}

Of the 64 cells, 18 have not been studied and 13 are structurally out of reach of the model class. Of the 33 cells with any evidence, 19 (58\,\%) rest on simulation alone and a further 6 on assumed parameters or IT telemetry, so 25 of 33 (76\,\%) involve no electrical measurement of the modeled object. Four cells reach laboratory or device tests, two reach measurements inside a facility or of unstated boundary, and two touch a grid-side record. The facility-level cells are the internal resonance measurements and the two comparisons of the composite load model with measured data-center sags, in which default parameters did not match and per-event refitting reduced the error~\cite{jenkins2024wecc,sleiman2025cmld}; because one record is at the switchboard and the other's boundary is not stated, we code that cell at T2. The grid-side cells hold a qualitative observation of constant-power behavior and a single-event replay with an imposed trip fraction. One cell therefore contains measured evidence \emph{against} a model, which is informative but is not evidence of fidelity. The percentages are descriptive statistics of this constructed matrix, not estimates over all data-center modeling studies. \emph{No cell holds a successful held-out validation of a frozen, facility-specific model.}

\begin{table}[!t]
\centering
\caption{The 33 evidence-bearing cells of Fig.~\ref{fig:matrix} by provenance of the reference (rows) and by what the comparison demonstrated (columns).}
\label{tab:provstatus}
\footnotesize
\setlength{\tabcolsep}{2.6pt}
\begin{tabular}{@{}l c c c c c c c@{}}
\toprule
Provenance & HV & MM & CA & QO & PA & MS & Total\\
\midrule
T1 grid-side record & 0 & 0 & 1 & 1 & 0 & 0 & 2\\
T2 facility-internal & 0 & 1 & 1 & 0 & 0 & 0 & 2\\
T3 laboratory/device & 0 & 0 & 4 & 0 & 0 & 0 & 4\\
T4/T2w parameters, telemetry & 0 & 0 & 0 & 0 & 6 & 0 & 6\\
T5 simulation & 0 & 0 & 0 & 0 & 0 & 19 & 19\\
\midrule
Total & 0 & 1 & 6 & 1 & 6 & 19 & 33\\
\bottomrule
\multicolumn{8}{@{}p{0.95\columnwidth}@{}}{\footnotesize HV successful held-out validation; MM independent comparison, mismatch; CA calibration or comparison without a held-out event; QO qualitative observation; PA parameterization; MS model-to-model or simulation only.}
\end{tabular}
\end{table}

Provenance and validation status are different dimensions, and Fig.~\ref{fig:pyramid} shows only the first. Table~\ref{tab:provstatus} separates them. A high-provenance reference does not imply a successful test: the facility-level cell with a measured comparison demonstrates a mismatch, and the grid-side cells hold a qualitative observation and a calibrated replay. A lower-provenance reference can still support a narrow claim, as the laboratory cells do for individual devices. The empty top of the pyramid is therefore a status, not a tier: no cell, at any provenance, holds a successful held-out validation.

Three patterns explain the shape of the matrix. First, the phenomena that new requirements emphasize, repeated dips and fast active-power recovery (Table~\ref{tab:reqs}), are exactly those that standard library models cannot represent or that have only been simulated. Second, most ``real data'' in the literature is IT telemetry: GPU utilization traces calibrate workload statistics but say nothing about how the power chain and protection translate them into POI current. Third, measurement inside facilities exists, but it concerns resonance and harmonics, not the disturbance response that drives load-loss events. Several papers also describe agreement between two models as ``validation'', or judge one model ``optimistic'' or ``realistic'' by comparison with another, unvalidated reference model~\cite{choi2026kiee}; we code such comparisons as simulation, because neither side is a measurement.

The matrix was coded by a single reviewer from the sources in Section~\ref{sec:method}. The coding is conservative in one direction: a cell is credited with the strongest evidence any source provides, even if most sources for that cell are weaker. The tier assignments involve judgment, for example coding a comparison whose measurement boundary is not stated~\cite{sleiman2025cmld} at T2, or a qualitative grid-side observation at T1. Plausible alternatives (T1 and T2 respectively) each move one cell between tiers but leave the 25 of 33 cells without electrical measurement unchanged, and the central finding does not depend on any tier assignment.

\section{Worked Cases}\label{sec:cases}
Table~\ref{tab:cases} applies the gates of Section~\ref{sec:framework} to eleven cases chosen because each isolates one way in which evidence and claim come apart. The release labels concern the stated claim, not the quality of the study.

\begin{table*}[!t]
\centering
\caption{Worked cases: evidence, the claim it is sometimes taken to support, the gate that decides, and the resulting release label.}
\label{tab:cases}
\footnotesize
\begin{tabularx}{\textwidth}{@{}>{\raggedright\arraybackslash}p{2.3cm} Y Y >{\raggedright\arraybackslash}p{2.0cm} >{\raggedright\arraybackslash}p{2.8cm}@{}}
\toprule
Case & What was measured or done & Claim at risk & Deciding gate & Release \\
\midrule
WECC data center~\cite{jenkins2024wecc} & 0.50\,pu sag at a $\sim$200\,MW site; switchboard current doubled; standard composite model compared & ``Standard library models represent data-center sag response'' & G5 (a real comparison) & Held: both measured comparisons~\cite{jenkins2024wecc,sleiman2025cmld} report a mismatch with this configuration\\
ERCOT Facility~L~\cite{yazdanpanah2026partial} & One partial-tripping event replayed in EMT; tripped fraction set to match; generic collector & ``The model is validated''; ``the site has Yg--Yg transformers'' & G4 (one event, fitted) & Conditional (calibration only, G4); winding inference conditional\\
PERC1~\cite{odunlami2026perc1,nerc_questionnaire} & Parameters from questionnaires; identifiability studied on synthetic PMU records & ``PERC1 with questionnaire values predicts ride-through'' & G4, G5 & Held: no measured comparison exists; one event could not identify all threshold--delay pairs\\
Dominion 14.7\,Hz~\cite{s3_147hz} & Months of synchrophasor data; point-on-wave check of aliasing; owner confirmed UPS input stage & ``Models can reproduce the oscillation'' & G2 (30\,Hz storage), G5 & Phenomenon and cause released at the measured boundary; model claim held (no model compared)\\
Same site, E3/E4~\cite{nerc_wp_disturbance} & 20\,\% vs 100\,\% load reduction at 0.72\,pu/67\,ms and 0.75\,pu/75\,ms & ``A static $V$--$t$ threshold model describes the site'' & G4, G5 (no model compared) & Held for the prediction: no model was compared with the records. The narrower observation, that two near-identical sags produced very different reductions, is released\\
Meta and Finnish facilities~\cite{sun2022p1,sun2022p2,alenius2020} & Switchboard and rack resonance; impedance measured at one rectifier and aggregated & ``Grid-level resonance risk is quantified'' & G2, G3 (boundary R/MSB$\rightarrow$POI; one rectifier aggregated) & Narrowed: internal mechanism released at MSB/R; POI claim conditional\\
Validated object $\neq$ claim~\cite{alkez2020,samanta2024emt,zamani2026esig} & Grid model calibrated to Irish frequency events; one miner PSU vs lab; PSU settings tuned to HIL & ``The data-center response model is validated'' & G3 & Narrowed to the component; facility claim held (no facility measurement)\\
Irish 204\,MW event~\cite{jimenezruiz2026} & Event cited as motivation; detailed model run on assumed parameters & ``The model reproduces observed behavior'' & G5 & Held for prediction; released as sensitivity study\\
Per-event CMLD refit~\cite{sleiman2025cmld} & Real sag waveforms (one phase); 133 parameters inverted by machine learning for each event; fault also tuned & ``The data-center CMLD is validated on other events'' & G4 (fit per event), G5 & Default configuration held (contradicted); tuned sets conditional (calibration, G4)\\
EirGrid demand cap~\cite{eirgrid2026framework} & Facility model data plus TSO simulations of credible faults; no grid testing & ``The aggregate data-center demand loss for a credible fault is known'' & G3, G5 & Held as a prediction (no measured comparison); usable only as a bound conditional on the submitted models\\
IESO model package~\cite{ieso2026lcl} & RMS models benchmarked against the facility EMT model; equipment type or commissioning tests & ``The submitted model represents the installed facility'' & G3 (two models agree; equipment, not model, tested) & Equipment compliance released narrowly; model fidelity held\\
\bottomrule
\end{tabularx}
\end{table*}

\begin{table*}[!t]
\centering
\caption{Gate outcomes for the worked cases of Table~\ref{tab:cases}. Claim type: O observation, S sensitivity, P prediction. \checkmark{} pass; $\times$ fail; C compared with an independent measurement and contradicted; -- not applicable to the claim type; $\cdot$ not assessed because an earlier gate already decides the label.}
\label{tab:gateout}
\footnotesize
\setlength{\tabcolsep}{3.5pt}
\begin{tabular}{@{}l >{\raggedright\arraybackslash}p{5.0cm} c c c c c c c c c >{\raggedright\arraybackslash}p{3.8cm}@{}}
\toprule
Case & Claim & Type & G0 & G1 & G2 & G3 & G4 & G5 & G6 & G7 & Label\\
\midrule
WECC data center & Standard library models represent data-center sag response & P & \checkmark & \checkmark & \checkmark & \checkmark & \checkmark & C & $\times$ & $\times$ & Held (contradicted)\\
ERCOT Facility L & The EMT facility model is validated & P & \checkmark & \checkmark & \checkmark & \checkmark & $\times$ & $\times$ & $\times$ & $\times$ & Conditional (calibration)\\
PERC1 & Questionnaire parameters predict ride-through & P & \checkmark & \checkmark & $\times$ & $\times$ & $\times$ & $\times$ & $\cdot$ & $\cdot$ & Held (no measured comparison)\\
Dominion 14.7\,Hz & The oscillation and its cause, at the measured boundary & O & \checkmark & \checkmark & \checkmark & \checkmark & -- & -- & \checkmark & -- & Released\\
Dominion 14.7\,Hz & Models reproduce the oscillation & P & \checkmark & \checkmark & \checkmark & $\times$ & $\cdot$ & $\times$ & $\cdot$ & $\cdot$ & Held (no model compared)\\
Same site, E3/E4 & Two near-identical sags gave 20\,\% and 100\,\% load reduction (cause not claimed) & O & \checkmark & \checkmark & \checkmark & \checkmark & -- & -- & \checkmark & -- & Released\\
Same site, E3/E4 & A static $V$--$t$ threshold model describes the site & P & \checkmark & \checkmark & \checkmark & \checkmark & $\cdot$ & $\times$ & $\cdot$ & $\cdot$ & Held (no model compared)\\
Meta and Finnish & Grid-level resonance risk is quantified & P & \checkmark & \checkmark & $\times$ & $\times$ & $\cdot$ & $\cdot$ & $\cdot$ & $\cdot$ & Narrowed to MSB/R\\
Validated object & The data-center response model is validated & P & \checkmark & \checkmark & $\times$ & $\times$ & $\cdot$ & $\cdot$ & $\cdot$ & $\cdot$ & Narrowed to the component; facility claim held\\
Irish 204\,MW & The model reproduces the observed behavior & P & \checkmark & \checkmark & $\times$ & $\times$ & $\times$ & $\times$ & $\cdot$ & $\cdot$ & Held (no measured comparison)\\
Irish 204\,MW & Results under the assumed facility behavior & S & \checkmark & \checkmark & -- & \checkmark & \checkmark & -- & -- & -- & Released as a sensitivity study\\
Per-event CMLD refit & The fitted CMLD is validated on other events & P & \checkmark & \checkmark & \checkmark & \checkmark & $\times$ & $\times$ & $\times$ & $\times$ & Conditional (calibration)\\
EirGrid demand cap & The demand loss for a credible fault is known & P & \checkmark & \checkmark & $\times$ & $\times$ & $\cdot$ & $\times$ & $\cdot$ & $\cdot$ & Held (no measured comparison)\\
IESO package & The submitted model represents the installed facility & P & \checkmark & \checkmark & $\times$ & $\times$ & $\cdot$ & $\cdot$ & $\cdot$ & $\cdot$ & Narrowed to equipment; model fidelity held\\
\bottomrule
\end{tabular}
\end{table*}

Four of the cases deserve comment. The WECC comparison is among the most informative results in the corpus because of its mismatch: it is one of only two instances in which a standard model was confronted with a measured data-center response, and in both the default configuration did not match~\cite{jenkins2024wecc,sleiman2025cmld}. Neither source states an acceptance criterion, so by gate G7 neither is a formal test in either direction; each reports a mismatch by its own judgment, and the 112\,\% error on the deep sag indicates a substantial disagreement. This shows that default parameters are inadequate for these facilities, not that the model structure cannot be made to fit. The second study then shows the natural next step and its limit: refitting each event separately fits each event well, but a set fitted to one event has not been tested on another. The ERCOT Facility~L study shows the opposite risk. Agreement with one event, obtained after setting the tripped fraction and fault impedance to match, was used to infer the facility's transformer connection. With one event and several free choices, other configurations may fit equally well; the inference is reasonable but conditional until an independent record or the owner confirms it, as happened in the 14.7\,Hz case. The two operator cases show the same gap at the level of policy. EirGrid's demand cap is a decision quantity computed entirely from submitted models, so its accuracy is that of the models; IESO's package tests the equipment and checks two models against each other, but nothing in it compares the model with the facility.

\section{Requirement-to-Model Mapping}\label{sec:mapping}
Table~\ref{tab:mapping} asks, for each feature of the requirements in Table~\ref{tab:reqs}, what a model must be able to represent, which model classes can represent it, and whether any model has been shown to reproduce it against measurement.

\begin{table*}[!t]
\centering
\caption{Requirement features, the model capability each needs, and the evidence that any model reproduces it.}
\label{tab:mapping}
\footnotesize
\begin{tabularx}{\textwidth}{@{}>{\raggedright\arraybackslash}p{3.0cm} >{\raggedright\arraybackslash}p{2.9cm} >{\raggedright\arraybackslash}p{1.8cm} >{\raggedright\arraybackslash}p{2.1cm} >{\raggedright\arraybackslash}p{1.6cm} Y@{}}
\toprule
Requirement feature & Model capability needed & CMLD elec. & PERC1 & EMT facility & Evidence a model reproduces it \\
\midrule
Single-dip voltage ride-through envelope & Voltage thresholds with time delays & Partial (no delays) & Yes & Yes & OEM parameters and simulation; both CMLD comparisons reported a mismatch~\cite{jenkins2024wecc,sleiman2025cmld}\\
Repeated dips, cumulative windows, reclosing~\cite{caiso_straw,spp_rt2026,ieeesa_ic25004} & Memory of preceding dips, counting logic & No & No (successor in development) & If logic is known & Simulation only; behavior observed in E2, E7--E9\\
Active-power floor during the sag~\cite{spp_rt2026,wang2026solvrt} & $P(V)$ during the sag, partial reduction & Constant $P$ & Constant $P$ until ceasing & Yes & Not studied against measurement\\
Recovery to 90\,\% in 0.5--3\,s~\cite{ercot_nogrr282,caiso_straw,spp_rt2026,eirgrid2026mpid345,aemc2026final} & Reconnection delay and ramp & Immediate partial restart & Delay and ramp & Yes & Typical settings (4--10\,s delay, 10--15\,s ramp) fall outside the requirement~\cite{kosterev2026perc1}\\
Overvoltage ride-through; post-transfer swell & Overvoltage-triggered transfer; site interaction & No & Not documented & If logic is known & Cascade observed in E18~\cite{wharton2026pjm}; no model\\
Compliance voltage definition (positive sequence vs worst phase)~\cite{epri2026rt} & Per-phase measurement and protection & No & No & Yes & Not studied\\
Unbalanced and partial tripping & Per-phase load behavior & No & No & Yes & One replay with imposed trip fraction~\cite{yazdanpanah2026partial}\\
\bottomrule
\end{tabularx}
\end{table*}

The mapping makes the gap concrete. The features that distinguish the new requirements from a simple envelope (memory of repeated dips, fast recovery, overvoltage response, and per-phase compliance) are those that the phasor models in current use either cannot represent or have never been compared with measurement for. EMT models can in principle represent all of them, but only if the facility's actual protection and control logic is known, and none has been validated against a measured event. A study that ``demonstrates compliance'' with these rules using any of these models therefore rests on assumptions about the facility rather than on evidence about it.

\section{Mitigations and the Evidence Behind Them}\label{sec:mitigation}
A growing literature proposes ways to make facilities compliant or to protect the grid from them. Table~\ref{tab:mitigation} groups the proposals and records what each was evaluated on. With two partial exceptions, rack-level hardware prototypes and a frequency-reserve pilot, every mitigation for disturbance response has been evaluated on a facility model whose own disturbance response is unvalidated. Sizing a grid-forming battery, a supercapacitor buffer or a braking resistor to keep a facility within an envelope therefore inherits the uncertainty of the facility model. This does not make the mitigation conclusions wrong, but it leaves their margin unknown: if the model misses dip counting or overvoltage-triggered transfer, the mitigation may be sized for the wrong event.

\begin{table*}[!t]
\centering
\caption{Proposed mitigations and the evidence on which they have been evaluated.}
\label{tab:mitigation}
\footnotesize
\begin{tabularx}{\textwidth}{@{}>{\raggedright\arraybackslash}p{3.4cm} Y Y l@{}}
\toprule
Mitigation & Purpose & Evaluated on & Tier\\
\midrule
Protection-setting changes~\cite{nerc_wp_disturbance,zechevikj2026,khomarudin2025} & Fewer transfers (count only long, deep dips; coordinate with reclosing) & Adopted after events; effect inferred from later events, not from a model test & T1 (practice)\\
Grid-forming BESS at the site~\cite{kundu2026,lyu2026,ross2026,azizi2025,azizi2026b,shamseldein2026b,mohammadi2026adapen,yogarathnam2026bess} & Smooth workload swings, damp oscillation, support ride-through & Simulated facility models, some driven by measured GPU power traces or scaled measured load profiles & T5 (T2w inputs)\\
Oscillation damping with BESS~\cite{azizi2026} & Damp workload-induced SSO & Simulation & T5\\
Supercapacitor power conditioning~\cite{seo2026} & Ride-through and fast support & Real-time simulation; hardware tests as future work & T5\\
Rack-level active power buffers~\cite{gu2026} & Absorb GPU transients near the load & Hardware prototype & T3\\
FACTS, E-STATCOM, HVDC~\cite{ilic2026,elliott2026sandia} & Voltage support, longer critical clearing time & Simulation; state-of-the-art surveys & T5\\
Breaker-switched braking resistors~\cite{ghosh2025} & Absorb surplus power after load loss & Theoretical demonstration & T5\\
Impedance shaping with co-located electrolysis~\cite{shamseldein2027} & Passivity and fault support & Simulation & T5\\
Workload modulation and UPS-based frequency response~\cite{tao2025,kim2025,sun2025,alapera2019,statnett2018} & Fast frequency response & Simulation; one UPS reserve pilot & T5; T1 (pilot)\\
Load modulation and GPU ramp limits~\cite{maheshwari2026} & Power quality & Simulation & T5\\
\bottomrule
\end{tabularx}
\end{table*}

The pattern matters because mitigations are where money is committed. A facility owner who installs a grid-forming battery to meet a 0.5\,s recovery deadline, or an operator who accepts that battery as making the facility compliant, is relying on two models at once: the battery's, which is typically validated as an inverter-based resource and whose own ride-through support depends on its current-priority setting and on grid strength~\cite{cheerla2026lvrt}, and the facility's, which is not. The first hardware-in-the-loop tests with physical power supplies and UPS units~\cite{zamani2026esig} are the natural place to test both together.

\section{Minimum Validation Protocol}\label{sec:protocol}
We propose the following minimum protocol for any claim that a data-center model predicts disturbance behavior relevant to an interconnection decision. It adapts the gates of Section~\ref{sec:framework} to what operators can realistically require.
\begin{enumerate}
\item \emph{State the context of use} $C$, including the decision and the requirement against which compliance is claimed (gate G0).
\item \emph{Record the right channels at the right boundary.} At the POI: three-phase point-on-wave voltage and current, with phasors at 120--240 frames per second rather than 30--60~\cite{nerc_trd,s3_147hz}. Inside the facility: UPS mode and transfer state, protection pickups and trips, and the active workload level, time-synchronized with the POI record~\cite{nerc_trd,albustami2026attribution}.
\item \emph{Assess identifiability, then calibrate.} State which parameters the available disturbances can determine. For PERC1 threshold--delay pairs, two disturbances with different voltage trajectories are the minimum shown to be informative~\cite{odunlami2026perc1}. The reason is structural: a single dip crosses each threshold at one depth and one duration, so it bounds a threshold--delay pair from one side only, and a second dip with a different trajectory is needed to bracket it; other parameters and model structures may need more events, or must be fixed from verified settings.
\item \emph{Freeze the model} (structure, parameters, firmware and protection settings) after calibration and before the comparison event, and record the version.
\item \emph{Test on a held-out event} of the disturbance class named in $C$: for ride-through claims, an event with multiple dips if the requirement counts dips.
\item \emph{Compare the endpoint named in the requirement} (for example, active power 2 or 3\,s after voltage recovery) with stated measurement and parameter uncertainty, and an acceptance criterion fixed in advance. The criterion must state the error normalization (for example, active-power error as a fraction of pre-disturbance power), the timing reference (for example, the instant the POI voltage first falls below 0.9\,pu), and the instrument: timing tolerances of a few milliseconds require point-on-wave or high-rate fault-recorder data, not PMU phasors reported at 30--60 frames per second. A developer's workshop proposal of $\pm$5\,\% in active power and $\pm$5\,ms in timing on benchmark events, which does not specify the event types~\cite{syed2026esig}, is a starting point, not an operator requirement.
\item \emph{Re-validate after material change} in UPS firmware, protection settings or IT equipment, as NERC's guidance already anticipates~\cite{nerc_trd}.
\end{enumerate}
The protocol is not new in kind: it is what inverter-based generators already supply~\cite{soco2025ibrmodel}, and Southern Company's performance-validation period~\cite{soco2026tcll} is a ready vehicle for steps 2, 5 and 6 if it adds an acceptance criterion. Until step~5 has been completed for at least one facility, compliance studies should be reported as conditional on the assumed facility logic, and sensitivity to that logic should be shown.

\emph{Choosing the metric.} The acceptance criterion of step~6 depends on the decision. Table~\ref{tab:metrics} lists candidate metrics by context of use. Tolerances should follow from the measurement uncertainty of the instrument and from the consequence of wrongly accepting the model (gate G7), and should be fixed before the held-out comparison; we do not propose universal numerical tolerances, and none of the values in the example below is a recommended standard.

\begin{table}[!t]
\centering
\caption{Candidate acceptance metrics by context of use. Tolerances are set per decision, from instrument uncertainty and the consequence of a wrong acceptance.}
\label{tab:metrics}
\footnotesize
\setlength{\tabcolsep}{3pt}
\begin{tabular}{@{}>{\raggedright\arraybackslash}p{2.2cm} >{\raggedright\arraybackslash}p{3.2cm} >{\raggedright\arraybackslash}p{2.45cm}@{}}
\toprule
Decision or claim & Metric & Data needed\\
\midrule
Ride-through (cease or stay) & Agreement of predicted and observed state per load group; error in the fraction of load that ceases & POI record with UPS and protection states\\
Active-power recovery & Error in recovery time; error in $P$ at the required instant, as a fraction of pre-disturbance $P$ & Point-on-wave or $\geq$120 frames/s\\
Trajectory fidelity & Normalized RMSE of $P(t)$ and $Q(t)$ over a stated window & Same as above\\
Current during the sag & Magnitude and phase error of current & Point-on-wave\\
Oscillations & Error in frequency, damping ratio and amplitude of the dominant modes & Sampling well above twice the highest mode frequency\\
Workload ramps & Error in ramp rate and in peak-to-peak power change; spectral content in fixed frequency bands & POI $P$ synchronized with workload telemetry\\
Parameter claims & Confidence or credible intervals of identified parameters & Several disturbances with different trajectories\\
Predictive reliability & Error distribution over all held-out events, not the best case & A set of held-out events of the claimed class\\
\bottomrule
\end{tabular}
\end{table}

\subsection{A synthetic worked example}\label{sec:example}
To show the protocol operating, rather than describe it, we apply it to synthetic data. Nothing in this example is measured: a known ``facility'' generates the records, so that the outcome of each step can be checked against the truth. The facility has three load paths. Two cease when the POI voltage stays below a threshold for a delay, reconnect after the voltage has exceeded 0.9\,pu for a set time and then ramp back, as in~\eqref{eq:dropout}--\eqref{eq:ramp}; the two paths carry 40\,\% and 35\,\% of the load, with thresholds of 0.70 and 0.60\,pu. A path that ceases twice within 30\,s stays on battery for 60\,s, a simple form of the dip memory in current requirements. The remaining 25\,\% rides through at constant power. The model to be validated has the same two-path structure without memory. Measurement noise of 1\,\% of pre-disturbance power is added to every record. Table~\ref{tab:exparams} specifies the experiment, and the scripts implement it.

The context of use is a ride-through and recovery claim at the POI for single and repeated dips, with active power over the 12\,s record as endpoint. Before any comparison we fixed the criterion: normalized RMSE of $P$ at most 5\,\% of pre-disturbance power, recovery-time error at most 0.5\,s, and error in the fraction of load that ceases at most 0.10. The values are illustrative only. Following steps~3 and~4, the model was calibrated on two single dips with different trajectories (A: 0.65\,pu for 0.15\,s; B: 0.55\,pu for 0.30\,s) and then frozen. The identifiability check flags a problem before any test. Parameter sets within 5\,\% of the best fit place the lower threshold anywhere between 0.55 and 0.65\,pu (true 0.60) and the upper one between 0.65 and the bound of 0.94\,pu (true 0.70), because no calibration dip falls in those bands; every other parameter is pinned closely (Table~\ref{tab:exparams}). The fitted lower threshold, 0.5500005\,pu, sits just above B's depth: with the strict inequality of~\eqref{eq:dropout} as implemented, it is the lowest value at which path~2 still ceases on B, an edge-of-data estimate that is itself the symptom.

\begin{figure}[!t]
\centering
\includegraphics[width=\columnwidth]{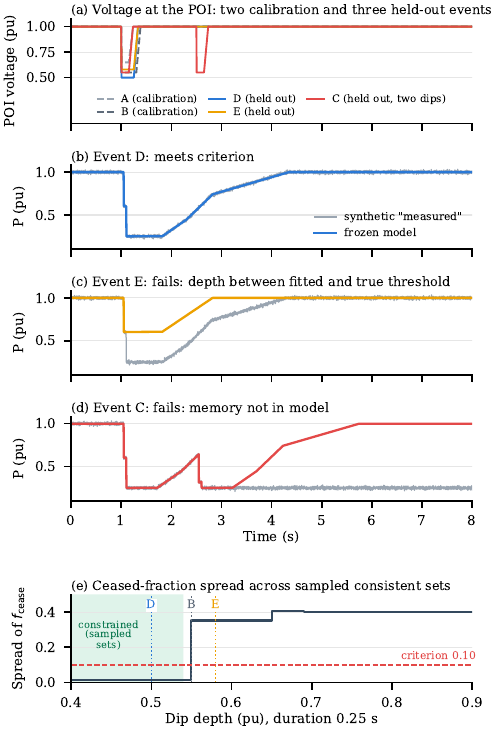}
\caption{Synthetic worked example of the minimum validation protocol. (a) POI voltage of the two calibration events (A, B) and three held-out events (D, E, C). (b)--(d) Synthetic ``measured'' active power and the frozen model on the held-out events; the first 8\,s of the 12\,s records are shown, and all metrics use the full record. (e) Preliminary predictive-identifiability check: spread of the predicted ceased fraction across the sampled parameter sets consistent with A and B (Table~\ref{tab:exparams}), for single dips of 0.25\,s; shading marks where the spread is within the 0.10 criterion. Recovery time and trajectory are not assessed in this panel. All data are synthetic.}
\label{fig:example}
\end{figure}

\begin{table}[!t]
\centering
\caption{Synthetic worked example: errors of the frozen model against the pre-set criterion (NRMSE $\leq$\,0.05, recovery-time error $\leq$\,0.5\,s, ceased-fraction error $\leq$\,0.10).}
\label{tab:example}
\footnotesize
\setlength{\tabcolsep}{2.2pt}
\begin{tabular}{@{}l l c c c c@{}}
\toprule
Event & Role & NRMSE & $\Delta t_{\mathrm{rec}}$ (s) & $\Delta f_{\mathrm{cease}}$ & Meets\\
\midrule
A: 0.65\,pu, 0.15\,s & Calibration & 0.010 & 0.04 & 0.01 & --\\
B: 0.55\,pu, 0.30\,s & Calibration & 0.011 & 0.05 & 0.01 & --\\
D: 0.50\,pu, 0.25\,s & Held out & 0.011 & 0.08 & 0.01 & Yes\\
E: 0.58\,pu, 0.25\,s & Held out & 0.139 & 1.27 & 0.36 & No\\
C: 2$\times$0.55\,pu & Held out & 0.591 & $\geq$6.85$^{a}$ & 0.01 & No\\
\bottomrule
\multicolumn{6}{@{}p{0.95\columnwidth}@{}}{\footnotesize $^{a}$Censored: the facility had not recovered when the record ended, so its recovery time is at least the remaining record length (9.35\,s); the value is a lower bound.}
\end{tabular}
\end{table}

Three held-out events then test the frozen model (Fig.~\ref{fig:example}, Table~\ref{tab:example}). Event D, a single dip outside the unidentified band, meets every criterion. Event E, a single dip at 0.58\,pu, inside the band, fails: the model keeps the 0.60\,pu path connected, predicts that 40\,\% rather than about 75\,\% of the load ceases, and recovers 1.3\,s too early. Event C, two dips 1.5\,s apart, fails by a wide margin: the facility latches on battery after the second dip, while the model, which has no memory, reconnects. The calibration fits were equally good in all three cases, so none of these outcomes could have been read from the calibration error.

Applied to this evidence, the gates give three labels. Event D shows event-specific agreement, so the claim is \emph{released} only for the conditions D tested: an isolated single dip of about 0.50\,pu lasting 0.25\,s. Whether that agreement extends to other dips is a G4 question. As a preliminary check, propagating the sampled calibration-consistent parameter sets of Table~\ref{tab:exparams} through single dips of 0.40--0.90\,pu and 0.05--0.40\,s gives predicted ceased fractions that agree within 0.015 for dips to 0.54\,pu or deeper, but spread by 0.35--0.41 between 0.56 and 0.90\,pu (Fig.~\ref{fig:example}e). This shows where a broader release could be sought; it does not establish one, because the sampled sets are not the full joint set of consistent parameters and the check covers the ceased fraction only, not recovery time or trajectory. A broader release would need a joint predictive-uncertainty assessment for every metric of the criterion and further held-out events. For shallower single dips the claim is \emph{held}: even the sampled sets disagree there, and the held-out comparison E contradicts the model inside that range. For repeated dips the claim is \emph{held}: the comparison C contradicts it, and in the terms of Fig.~\ref{fig:matrix} the model structure cannot represent the memory (X). A compliance claim against a requirement that counts dips would therefore be held, even though the same model fits both calibration events to within about 1\,\%. The example contains no information about any real facility; it shows only how the gates convert held-out results into a label. The simulation and figure scripts are available from the corresponding author on request.

\begin{table}[!t]
\centering
\caption{Specification of the synthetic example; the supplied scripts implement it. Ranges are parameter values for which the calibration loss stays within 5\,\% of its minimum, from a one-at-a-time profile over $\pm$60\,\% of each fitted value (a lower bound on the consistent set).}
\label{tab:exparams}
\footnotesize
\setlength{\tabcolsep}{3pt}
\begin{tabular}{@{}l c c c@{}}
\toprule
Parameter & True & Fitted & Consistent range\\
\midrule
$f_1$ & 0.40 & 0.398 & 0.398--0.406\\
$V_{d1}$ (pu) & 0.70 & 0.681 & 0.654--0.94$^{a}$\\
$T_{d1}$ (s) & 0.050 & 0.049 & 0.049--0.050\\
$T_{r1}$ (s) & 0.50 & 0.504 & 0.484--0.514\\
$R_1$ (pu/s) & 1.0 & 0.997 & 0.977--1.017\\
$f_2$ & 0.35 & 0.352 & 0.345--0.359\\
$V_{d2}$ (pu) & 0.60 & 0.550 & 0.550--0.649\\
$T_{d2}$ (s) & 0.100 & 0.102 & 0.098--0.102\\
$T_{r2}$ (s) & 1.00 & 0.970 & 0.931--1.009\\
$R_2$ (pu/s) & 0.50 & 0.487 & 0.478--0.507\\
\midrule
\multicolumn{4}{@{}p{0.95\columnwidth}@{}}{\footnotesize \emph{Simulation.} Time step 2\,ms; 12\,s records; dips start at 1.0\,s (C: 1.0 and 2.5\,s) with an 80\,ms linear voltage recovery; a path ceases when $V<V_d$ for $T_d$ and reconnects after $V>0.9$\,pu for $T_r$, then ramps at $R$; memory (truth only): two ceases within 30\,s latch the path off for 60\,s; 25\,\% constant power. Noise: Gaussian, $\sigma=0.01$\,pu, seed 7. \emph{Fit.} Objective: sum over A and B of the mean squared error of $P$ over the full record; Nelder--Mead from six starts ($\pm$20\,\% about an initial guess, seeds 0--5); constraints $0.3<V_{d2}<V_{d1}<0.95$, $f_1+f_2\le1$, $R>0.05$. \emph{Metrics.} NRMSE of $P$ over the full record, normalized by $P_0$; ceased fraction $1-\min P$ of a 50\,ms moving mean within 3\,s of dip onset; recovery time from clearance of the last dip to the last sample below $0.9P_0$, censored at the record end. $^{a}$Upper constraint.}\\
\bottomrule
\end{tabular}
\end{table}

\section{Relation to Existing Reviews}\label{sec:positioning}
Table~\ref{tab:positioning} compares this review with the reviews and syntheses closest to it. Several already note that data-center models need validation against measured data~\cite{viththarachchige2026,ghazavi2026pem}, and three grade evidence maturity in adjacent domains: sustainability claims~\cite{gilani2026energies}, the demonstrated scale of asset and threat behaviors~\cite{shrestha2026esm}, and flexibility services~\cite{shihab2027apen}. The review closest in scope, a 2026 preprint on large computational load modeling, is narrative: it reports no systematic search or evidence grading, and the validation of its own data-center EMT models is a comparison with corresponding PSS/E models~\cite{javadi2026review}, which in our terms is model-to-model. Reviews of AI-based load modelling~\cite{singh2025aireview} and of data-driven converter modeling~\cite{subedi2025overview} survey methods without auditing what the methods were validated against. We do not claim that grading evidence is new. To our knowledge, however, no prior review audits the evidence behind \emph{dynamic disturbance-response models} of data centers against measured events, applies decision gates that separate calibration from validation and validated object from claimed object, or maps current ride-through requirements to what the models in use can represent.

\begin{table*}[!t]
\centering
\caption{Positioning relative to related reviews and syntheses (2025--2026).}
\label{tab:positioning}
\footnotesize
\begin{tabularx}{\textwidth}{@{}>{\raggedright\arraybackslash}p{4.4cm} Y >{\raggedright\arraybackslash}p{3.6cm} >{\raggedright\arraybackslash}p{2.0cm} >{\raggedright\arraybackslash}p{2.0cm}@{}}
\toprule
Review & Primary scope & Grades evidence? & Audits model validation vs. events & Maps requirements to models\\
\midrule
Gyang \emph{et al.}~\cite{gyang2026ojies} & Architectures, component models, interconnection requirements, flexibility & No & No & Partly (lists requirements)\\
Viththarachchige \emph{et al.}~\cite{viththarachchige2026} & Load-model critique; model-domain selection & No (calls for field data) & No & No\\
Gilani \emph{et al.}~\cite{gilani2026energies} & Sustainable integration (energy, carbon, cooling) & Yes: reporting quality of quantitative claims & No & No\\
Dhakal \emph{et al.}~\cite{dhakal2026review} & Planning to feeder assets; flexibility & Source type only & No & No\\
Shrestha \emph{et al.}~\cite{shrestha2026esm} & Asset vs threat identity across time scales & Yes: scale and control of demonstrations & No & No\\
Shihab \emph{et al.}~\cite{shihab2027apen} & Grid-interactive operation and flexibility & Yes: evidence maturity / TRL of flexibility & No & No\\
Ghazavi Dozein \emph{et al.}~\cite{ghazavi2026pem} & Large-load modeling practice (ESIG task force) & No (recommends validation procedures) & No & Partly\\
Xie \emph{et al.}~\cite{xie2026procieee} & Reliability and affordability under uncertainty and variation & No & No & No\\
Javadi \emph{et al.}~\cite{javadi2026review} (preprint) & Large-load modeling for transmission planning: ZIP to PERC1, LDEV1 and EMT; utility practice & No & No (EMT model checked against PSS/E model) & Partly (reviews requirements)\\
Singh \emph{et al.}~\cite{singh2025aireview}; Subedi \emph{et al.}~\cite{subedi2025overview} & AI-based load modelling; data-driven converter modeling & No & No & No\\
Chen \emph{et al.}~\cite{chen2026nexus}; Mughees \emph{et al.}~\cite{mughees2026review}; Song \emph{et al.}~\cite{song2026ojpel}; Ginzburg-Ganz \emph{et al.}~\cite{ginzburg2026energies}; Seshmasetti \emph{et al.}~\cite{seshmasetti2025isgt}; Mohammadi \emph{et al.}~\cite{mohammadi2026adapen}; Maheshwari \emph{et al.}~\cite{maheshwari2026}; Parikh \emph{et al.}~\cite{parikh2025}; Wang \emph{et al.}~\cite{wang2026} & Demand, grid impacts, power architecture, storage, power quality & No & No & No\\
\textbf{This review} & Credibility of dynamic data-center load models & Yes: reference object and validated object & Yes & Yes\\
\bottomrule
\end{tabularx}
\end{table*}

\section{Research Agenda}\label{sec:agenda}
The findings point to ten concrete needs.
\begin{enumerate}
\item \emph{A shared disturbance library.} Anonymized POI point-on-wave records of data-center disturbance responses, with UPS and protection states, would allow the first held-out validations.
\item \emph{A validation benchmark for PERC1 and its successor}: identification on two disturbances with different voltage slopes and testing on a third, following the identifiability result of~\cite{odunlami2026perc1}.
\item \emph{Models with memory and overvoltage response}, able to represent dip counting, cumulative windows and the overvoltage-triggered transfers seen on 22 July 2026.
\item \emph{Aggregate validation.} Site-by-site models must be checked against multi-site events, because sites interact through network voltage.
\item \emph{Linking IT telemetry to POI response.} Joint datasets of workload and POI measurement are needed to test whether workload-calibrated models predict electrical behavior~\cite{albustami2026attribution}.
\item \emph{Measurement and certification standards}: a common compliance-voltage definition, minimum recorder bandwidth, and an equipment-level test framework that maps immunity standards to grid requirements~\cite{epri2026rt}.
\item \emph{Mitigations tested against validated plants.} Grid-forming storage, braking resistors, supercapacitors and UPS controls have so far been evaluated only on unvalidated facility models; hardware-in-the-loop tests with physical PSUs and UPS units~\cite{zamani2026esig} are a practical intermediate step.
\item \emph{Acceptance criteria for load models.} Operators that already monitor facilities after commissioning~\cite{soco2026tcll} could publish the criterion by which a load model is accepted, as they do for generators~\cite{soco2025ibrmodel}.
\item \emph{Data-driven models tested on withheld measured events.} Learned and hybrid models should be trained on some recorded data-center disturbances and tested on others against an open benchmark; public sub-second AI-training power data now cover the workload side~\cite{elsayed2026dataset}, but no comparable dataset of electrical disturbance responses exists.
\item \emph{Precise reporting language.} Studies should label each comparison as calibration, model-to-model agreement or held-out validation, and name the object validated.
\end{enumerate}

\section{Limitations of This Review}\label{sec:limitations}
Four limitations bound these findings. First, the review covers the published and publicly posted record; utilities and vendors may hold unpublished comparisons of facility models with measured events, and the central finding should be read as ``none published'', which is itself the gap a planner relying on public evidence faces. Second, coding was performed by one reviewer; the full coding is available on request so that others can check or recode it. Third, requirements are changing month by month (their status was last checked on 19 November 2026), several of the values in Table~\ref{tab:reqs} come from drafts, and some are taken from a secondary survey or read from figures, as marked. Fourth, event magnitudes are as reported by operators and are approximate.

\section{Conclusions}\label{sec:conclusion}
Grid operators now require dynamic models of large computational loads and use them to demonstrate compliance with new ride-through rules. Across 160 documents, both comparisons of a standard library model with measured data-center events reported that the default model did not match, and the one refit was made event by event, so no fitted set was tested on an event it had not seen. Within the publicly accessible literature and operator documents reviewed, no study demonstrated a successful held-out validation of a frozen, facility-specific data-center model against a predefined acceptance criterion; this is a gap in accessible evidence, not proof that no such validation has been done. Three quarters of the model--phenomenon combinations with any evidence rest on simulation, assumed parameters or IT telemetry. The features that distinguish the new requirements (memory of repeated dips, recovery within 0.5--3\,s, proportional power reduction during the sag, overvoltage response) are those the phasor models in use cannot represent or have never been tested on, yet the operator procedures examined rely on comparisons with other models and, at most, an unspecified check against monitored data, with no predefined acceptance criterion, and proposed mitigations have, with minor exceptions, been tested only on unvalidated facility models. An update search on data-driven and AI-based models (70 studies) does not change this: learned models are now tested on held-out events, but only simulated ones, and none of the field records they use is of a data center.

None of this makes current models useless: they remain appropriate for sensitivity studies and for exploring what facility logic would be needed to comply. It does mean that compliance claims should be reported as conditional on assumed facility behavior until models are tested against held-out measurements. The context-of-use gates, release labels and minimum protocol proposed here are intended to make that distinction explicit, and the disturbance records that utilities have already collected are sufficient to begin.

\section*{Acknowledgment}
During manuscript preparation, the authors used Anthropic Claude for writing assistance. The authors independently verified all data, analyses, quotations, references, and substantive claims, edited the output as needed, and take full responsibility for the final content.

\section*{Data Availability}
The per-source coding, the derivation scripts and the worked-example scripts supporting the findings of this review are available from the corresponding author on reasonable request.

\section*{Appendix A: Coding Rules}
Table~\ref{tab:codebook} gives the rules used to code each cell of the evidence matrix, so that the coding can be checked or repeated.

\begin{table}[!h]
\centering
\caption{Coding rules for the evidence matrix (Fig.~\ref{fig:matrix}).}
\label{tab:codebook}
\scriptsize
\renewcommand{\arraystretch}{1.12}
\begin{tabularx}{\columnwidth}{@{}l Y@{}}
\toprule
Code & Assigned when the strongest source for the cell\ldots\\
\midrule
N & addresses the phenomenon with no model of this class\\
X & cannot be addressed because the model structure cannot express it (e.g., memory, per-phase behavior)\\
S & compares the model only with simulation or another model\\
P & sets parameters from OEM data or questionnaires, or calibrates to IT telemetry\\
L & compares a device or emulator model with laboratory measurements\\
F & compares with measurements inside a facility (switchboard, rack)\\
E & reproduces one grid event after fitting at least one quantity to it\\
M & compares with a grid event and reports a mismatch with default parameters\\
Q & observes the behavior qualitatively in grid records, without a model comparison\\
\midrule
\multicolumn{2}{@{}p{0.95\columnwidth}@{}}{\emph{Scope.} Telemetry or parameters count as P only for the phenomena they describe (e.g., IT traces inform ramps, not ride-through); Q applies only where the source itself uses grid records to assess an assumption of the class; frequency-support services, faults internal to the facility and steady-state power-quality models are outside the eight phenomena and recorded as such. \emph{Precedence.} A cell takes the code of the source with the highest-provenance measurement (grid-side $>$ facility-internal or unstated boundary $>$ laboratory $>$ parameters or telemetry $>$ simulation), whether that source supports or contradicts the model; Q, E and M are tiered by their measurement boundary. \emph{Unknown boundary.} A measured record whose boundary is not stated is coded one tier below grid-side (T2). \emph{Tiers.} S = T5; P = T4/T2w; L = T3; F and M = T2 (in this matrix); E and Q = T1. Codes record provenance; whether a comparison succeeded is recorded separately (Table~\ref{tab:provstatus}), where a successful held-out validation of a frozen, facility-specific model would appear as a validation outcome; none was found. \emph{Overvoltage.} Overvoltage ride-through is coded within the VRT and cease--reconnect column, not as a separate phenomenon. \emph{Overlapping classes.} The class is that of the model a study actually runs, so a study running two models is coded under both; six of the 75 matrix-coded studies enter two classes. Seven cells rest only on such sources, all coded S or P; coding them N instead would leave 18 of 26 evidence-bearing cells without electrical measurement and no held-out validation, so the findings do not depend on the overlap. \emph{Conflicting sources.} Where sources at the top tier of a cell disagree, the cell takes the contradicting code (M) and the register keeps both; in this corpus no cell had co-ranked sources with different codes. The source-to-cell register is available on request.}\\
\bottomrule
\end{tabularx}
\end{table}

\section*{Appendix B: Cell-to-Source Map}
Table~\ref{tab:cellmap} lists every evidence-bearing cell of Fig.~\ref{fig:matrix} with its code, tier and the sources that set it; the 31 cells coded N or X have no supporting source by definition. The per-source coding of all 160 documents, with scripts that derive every cell by the precedence rule of Table~\ref{tab:codebook} and regenerate every count in Section~\ref{sec:matrix}, is available from the corresponding author on reasonable request.

\begin{table}[!h]
\centering
\caption{Cell-to-source map for the 33 evidence-bearing cells. Sources listed are those that establish the cell's tier; where several studies tie at that tier, the complete set of co-ranked sources is available on request.}
\label{tab:cellmap}
\scriptsize
\renewcommand{\arraystretch}{1.12}
\begin{tabularx}{\columnwidth}{@{}>{\raggedright\arraybackslash}p{1.6cm} Y l l >{\raggedright\arraybackslash}p{1.5cm}@{}}
\toprule
Model class & Phenomenon & Code & Tier & Sources\\
\midrule
Static / ZIP & VRT /\allowbreak  cease-reconnect & Q & T1 & \cite{nerc_trd,nerc_wp_disturbance}\\
Static / ZIP & Ramps /\allowbreak  workload variability & S & T5 & \cite{reyes2024}\\
CMLD electronic & VRT /\allowbreak  cease-reconnect & M & T2 & \cite{jenkins2024wecc,sleiman2025cmld}\\
CMLD electronic & Active-power recovery & S & T5 & \cite{choi2026kiee}\\
CMLD electronic & Frequency ride-through & S & T5 & \cite{choi2026kiee}\\
CMLD electronic & Forced /\allowbreak  sub-sync. oscillation & S & T5 & \cite{nerc_trd}\\
PERC1 & VRT /\allowbreak  cease-reconnect & P & T4/T2w & \cite{kosterev2026perc1,nerc_questionnaire}\\
PERC1 & Active-power recovery & P & T4/T2w & \cite{kosterev2026perc1}\\
PERC1 & Frequency ride-through & P & T4/T2w & \cite{kosterev2026perc1}\\
PERC1 & Forced /\allowbreak  sub-sync. oscillation & S & T5 & \cite{pnnl_emt}\\
Component-level phasor & VRT /\allowbreak  cease-reconnect & S & T5 & \cite{jimenezruiz2026}\\
Component-level phasor & Repeated dips /\allowbreak  reclosing & S & T5 & \cite{zechevikj2026}\\
Component-level phasor & Active-power recovery & S & T5 & \cite{jimenezruiz2026}\\
Component-level phasor & Frequency ride-through & S & T5 & \cite{gyang2026tpwrs}\\
Component-level phasor & Ramps /\allowbreak  workload variability & P & T4/T2w & \cite{lu2026}\\
Component-level phasor & Forced /\allowbreak  sub-sync. oscillation & S & T5 & \cite{arifin2026}\\
Workload-calibrated & VRT /\allowbreak  cease-reconnect & S & T5 & \cite{lu2026pattern}\\
Workload-calibrated & Ramps /\allowbreak  workload variability & P & T4/T2w & \cite{lu2026pattern,sreejith2026,chaudhary2026semimarkov}\\
Workload-calibrated & Forced /\allowbreak  sub-sync. oscillation & P & T4/T2w & \cite{xie2026procieee,ko2026wide}\\
Impedance (PSU/UPS) & Forced /\allowbreak  sub-sync. oscillation & S & T5 & \cite{shen2025,barati2026}\\
Impedance (PSU/UPS) & Resonance /\allowbreak  harmonics & F & T2 & \cite{sun2022p1,sun2022p2,alenius2020}\\
EMT facility & VRT /\allowbreak  cease-reconnect & S & T5 & \cite{ercot_pscad,pnnl_emt}\\
EMT facility & Repeated dips /\allowbreak  reclosing & S & T5 & \cite{zechevikj2026}\\
EMT facility & Active-power recovery & S & T5 & \cite{wang2026solvrt}\\
EMT facility & Frequency ride-through & S & T5 & \cite{pnnl_emt}\\
EMT facility & Ramps /\allowbreak  workload variability & S & T5 & \cite{lyu2026}\\
EMT facility & Forced /\allowbreak  sub-sync. oscillation & S & T5 & \cite{azizi2026}\\
EMT facility & Resonance /\allowbreak  harmonics & S & T5 & \cite{afzaal2025}\\
EMT facility & Unbalanced /\allowbreak  partial tripping & E & T1 & \cite{yazdanpanah2026partial}\\
EMT device / emulator & VRT /\allowbreak  cease-reconnect & L & T3 & \cite{samanta2024emt,zamani2026esig}\\
EMT device / emulator & Active-power recovery & L & T3 & \cite{sun2022,sun2021}\\
EMT device / emulator & Ramps /\allowbreak  workload variability & L & T3 & \cite{gu2026,sun2022}\\
EMT device / emulator & Resonance /\allowbreak  harmonics & L & T3 & \cite{sun2019pfc,zhu2023}\\
\bottomrule
\end{tabularx}
\end{table}

\bibliographystyle{IEEEtran}
\bibliography{refs,refs_auto,refs_dd}

@techreport{nerc_ir2025,
  author      = {{North American Electric Reliability Corporation}},
  title       = {Incident Review: Considering Simultaneous Voltage-Sensitive Load Reductions},
  institution = {NERC},
  address     = {Atlanta, GA, USA},
  month       = jan,
  year        = {2025},
  url         = {https://www.nerc.com/globalassets/our-work/reports/event-reports/incident_review_large_load_loss.pdf}
}

@techreport{nerc_wp_disturbance,
  author      = {{NERC Large Loads Working Group}},
  title       = {Large Load Disturbance Performance -- Impact Analysis and Ride-Through Recommendations},
  institution = {NERC},
  type        = {Draft White Paper},
  year        = {2026}
}

@techreport{nerc_l3alert,
  author      = {{North American Electric Reliability Corporation}},
  title       = {Level 3 Alert -- Essential Actions to Industry: Computational Load Modeling, Studies, Instrumentation, Commissioning, Operations, Protection, and Control},
  institution = {NERC},
  month       = may,
  year        = {2026},
  url         = {https://www.nerc.com/globalassets/programs/bpsa/alerts/level-3-computational-load-alert.pdf}
}

@techreport{nerc_trd,
  author      = {{NERC Load Modeling Working Group}},
  title       = {Data Center Load Modeling Technical Reference},
  institution = {North American Electric Reliability Corporation},
  address     = {Atlanta, GA},
  type        = {Technical Reference (final)},
  month       = jul,
  year        = {2026},
  note        = {Dated 30 Jul. 2026; posted Aug. 2026}
}

@misc{ercot_nogrr282,
  author       = {P. Gravois},
  title        = {Large Electronic Load Ride-Through Requirements: {NOGRR} 282 and {NPRR} 1308},
  howpublished = {ERCOT System Protection Working Group presentation},
  month        = feb,
  year         = {2026}
}

@techreport{caiso_straw,
  author      = {{California ISO Large Load Technical Requirements Working Group}},
  title       = {Large Load Technical Requirements -- Straw Proposal},
  institution = {California ISO},
  month       = jun,
  year        = {2026},
  url         = {https://stakeholdercenter.caiso.com/InitiativeDocuments/Large-Load-Technical-Requirements-Straw-Proposal-Jun-15-2026.pdf}
}

@misc{kosterev2026perc1,
  author       = {D. Kosterev},
  title        = {Using {PERC1} Model for Modeling Large Data Centers},
  howpublished = {Presentation, WECC Modeling and Validation Subcommittee},
  month        = may,
  year         = {2026}
}

@techreport{ercot_pscad,
  author      = {X. Wang and X. Chen and P. Enjeti and F. C. Lin and A. Yazdanpanah and J. Rose and Y. Cheng and A. Quedan and P. Kansal},
  title       = {Dynamic Modeling of {AI} Data Center Load in {PSCAD}, Version 1.0},
  institution = {Texas A\&M University and ERCOT},
  year        = {2026},
  url         = {https://www.ercot.com/about/grit/large-load-modeling}
}

@inproceedings{mishra2026discerning,
  author    = {C. Mishra and B. Pudasaini and K. A. Vance and T. J. Purcell and J. {Delaree Jr.} and K. D. Jones},
  title     = {Discerning Real-World Data Center Load Behavior and the Need for Long Term Data Analysis},
  booktitle = {Proc. IEEE PES General Meeting},
  year      = {2026},
  doi       = {10.1109/PESGM58988.2026.11693766}
}

@article{sun2022p1,
  author  = {J. Sun and M. Xu and M. Cespedes and M. Kauffman},
  title   = {Data Center Power System Stability -- {Part I}: Power Supply Impedance Modeling},
  journal = {CSEE J. Power Energy Syst.},
  volume  = {8},
  number  = {2},
  pages   = {403--419},
  year    = {2022},
  doi     = {10.17775/CSEEJPES.2021.02010}
}

@article{sun2022p2,
  author  = {J. Sun and M. Mihret and M. Cespedes and D. Wong and M. Kauffman},
  title   = {Data Center Power System Stability -- {Part II}: System Modeling and Analysis},
  journal = {CSEE J. Power Energy Syst.},
  volume  = {8},
  number  = {2},
  pages   = {420--438},
  year    = {2022},
  doi     = {10.17775/CSEEJPES.2021.02020}
}

@techreport{statnett2018,
  author      = {{Statnett SF}},
  title       = {Fast Frequency Reserves 2018 -- Pilot for raske frekvensreserver},
  institution = {Statnett},
  address     = {Oslo, Norway},
  year        = {2018}
}

@inproceedings{alapera2019,
  author    = {I. Alaper\"{a} and J. Paananen and K. Dalen and K. Hornnes and S. Honkapuro},
  title     = {Fast Frequency Response from a {UPS} System of a Data Center, Background, and Pilot Results},
  booktitle = {Proc. 16th Int. Conf. European Energy Market (EEM)},
  year      = {2019},
  doi       = {10.1109/EEM.2019.8916344}
}

@inproceedings{samanta2024emt,
  author    = {A. Samanta and S. Majumder and H. Ibrahim and P. Enjeti and L. Xie},
  title     = {Electromagnetic Transient Model of Cryptocurrency Mining Loads for Low-Voltage Ride Through Assessment in Transmission Grids},
  booktitle = {Proc. IEEE PES General Meeting},
  year      = {2024},
  doi       = {10.1109/PESGM51994.2024.10689224}
}

@article{jimenezruiz2026,
  author  = {A. Jim\'{e}nez-Ruiz and F. Milano},
  title   = {Data Centre Model for Transient Stability Analysis of Power Systems},
  journal = {IET Gener. Transm. Distrib.},
  year    = {2026},
  doi     = {10.1049/gtd2.70381},
  note    = {Preprint arXiv:2505.16575}
}

@techreport{pnnl_emt,
  author      = {B. A. Ross and S. Nekkalapu and X. Lyu and J. Follum},
  title       = {Electromagnetic Transient Modeling of Large Data Centers for Grid-Level Studies},
  institution = {Pacific Northwest National Laboratory},
  number      = {PNNL-38817 (Beta)},
  month       = aug,
  year        = {2026},
  url         = {https://www.pnnl.gov/main/publications/external/technical_reports/PNNL-38817beta.pdf}
}

@misc{odunlami2026perc1,
  author       = {B. G. Odunlami and M. Netto},
  title        = {A Hybrid Systems Approach to Parameterizing the {PERC1} Model for Data Centers Using Synchrophasor Measurements},
  howpublished = {arXiv:2611.21144},
  year         = {2026}
}

@misc{albustami2026attribution,
  author       = {A. A. Albustami and A. F. Taha and M. Alwahdanee and Z. S. Muhammad},
  title        = {Data Center Load Reductions: Attribution, Measurement Limits, and Estimation},
  howpublished = {arXiv:2610.38216},
  year         = {2026}
}

@inproceedings{yazdanpanah2026partial,
  author    = {A. Yazdanpanah and P. Gravois and J. Rose and M. Rahmatian and J. Zong},
  title     = {Impact of Unbalanced Partial Tripping of Large Electronic Loads on Local Power Quality},
  booktitle = {Proc. IEEE PES General Meeting},
  year      = {2026},
  doi       = {10.1109/PESGM58988.2026.11693488}
}

@inproceedings{spp_ops2026,
  author    = {{Southwest Power Pool} and others},
  title     = {Integration and Operational Considerations of Large Loads in {Southwest Power Pool}},
  booktitle = {Proc. IEEE PES General Meeting},
  year      = {2026},
  doi       = {10.1109/PESGM58988.2026.11693181}
}

@inproceedings{spp_rt2026,
  author    = {M. Sedighizadeh and D. Baker},
  title     = {Ride-Through and Delivery Point Study Requirements for High Impact Large Loads in {Southwest Power Pool}},
  booktitle = {Proc. IEEE PES General Meeting},
  year      = {2026},
  doi       = {10.1109/PESGM58988.2026.11693347}
}

@article{entergy2026,
  author  = {M. Peterson and K. McCulla and S. Kolluri and O. Aluko},
  title   = {Wired for Growth: Planning for Data Centers at {Entergy}},
  journal = {IEEE Power Energy Mag.},
  year    = {2026},
  doi     = {10.1109/MPE.2025.3638729}
}

@article{gyang2026tpwrs,
  author  = {P. P. Gyang and P. Chakraborty and L. Meegahapola and X. Yu},
  title   = {Dynamic Modeling of a Data Center for Power System Stability Studies},
  journal = {IEEE Trans. Power Syst.},
  volume  = {41},
  number  = {3},
  pages   = {2317--2332},
  doi     = {10.1109/TPWRS.2025.3639163},
  month   = may,
  year    = {2026},
  doi     = {10.1109/TPWRS.2025.3639163}
}

@misc{lu2026pattern,
  author       = {S. Lu and C. Xiao and Y. Weng},
  title        = {Dynamic Load Model for Data Centers with Pattern-Consistent Calibration},
  howpublished = {arXiv:2602.07859},
  year         = {2026}
}

@article{s3_147hz,
  author  = {C. Mishra and L. Vanfretti and J. {Delaree Jr.} and T. J. Purcell and K. D. Jones},
  title   = {Understanding the Inception of 14.7 {Hz} Oscillations Emerging from a Data Center},
  journal = {Sustainable Energy, Grids and Networks},
  year    = {2025},
  doi     = {10.1016/j.segan.2025.101735}
}

@standard{asme_vv40,
  organization = {ASME},
  title        = {Assessing Credibility of Computational Modeling through Verification and Validation: Application to Medical Devices},
  number       = {V\&V 40-2018},
  year         = {2018}
}

@misc{wharton2026pjm,
  author       = {M. Wharton},
  title        = {July 22, 2026 {Dominion} Load Transfer Event},
  howpublished = {PJM Operating Committee, Item 05},
  month        = aug,
  year         = {2026}
}

@article{alenius2020,
  author  = {H. Alenius and T. Roinila},
  title   = {Impedance-Based Stability Analysis of Paralleled Grid-Connected Rectifiers: Experimental Case Study in a Data Center},
  journal = {Energies},
  volume  = {13},
  number  = {8},
  pages   = {2109},
  year    = {2020},
  doi     = {10.3390/en13082109}
}

@techreport{ieeesa_ic25004,
  author      = {{IEEE SA Industry Connections}},
  title       = {Review of Industry Efforts and Standards: Data Center Design for {IT} Load Protection},
  institution = {IEEE Standards Association},
  number      = {IC25-004},
  year        = {2026}
}

@misc{jenkins2024wecc,
  author       = {S. Jenkins},
  title        = {Datacenter Load Modeling},
  howpublished = {Presentation, Western Electricity Coordinating Council},
  month        = jan,
  year         = {2024}
}

@misc{zamani2026esig,
  author       = {A. Zamani and A. Ghorashi},
  title        = {Considerations for {EMT} Modeling of Data Center Loads},
  howpublished = {ESIG Spring Technical Workshop},
  month        = mar,
  year         = {2026}
}

@misc{nerc_questionnaire,
  author       = {{NERC Load Modeling Working Group}},
  title        = {Data Center Information Collection Questionnaire},
  howpublished = {North American Electric Reliability Corporation},
  month        = mar,
  year         = {2026}
}

@article{alkez2020,
  author  = {D. {Al Kez} and others},
  title   = {Manipulation of Static and Dynamic Data Center Power Responses to Support Grid Operations},
  journal = {IEEE Access},
  volume  = {8},
  year    = {2020},
  doi     = {10.1109/ACCESS.2020.3028548}
}

@inproceedings{xie2026eenergy,
  author    = {Y. Xie and W. Cui and A. Wierman},
  title     = {Data Center Voltage Ride-Through: Emerging Challenges and Opportunities},
  booktitle = {Proc. 17th ACM Int. Conf. Future and Sustainable Energy Systems (e-Energy)},
  pages     = {379--385},
  year      = {2026},
  doi       = {10.1145/3744255.3798125}
}

@article{choi2026kiee,
  author  = {Y.-B. Choi and H.-J. Jo and W.-Y. Kim and M. Kim and S. Kim},
  title   = {Transient Stability Impact Comparison of Data Center Load Models},
  journal = {Trans. Korean Inst. Electr. Eng.},
  volume  = {75},
  number  = {7},
  pages   = {1459--1471},
  year    = {2026},
  doi     = {10.5370/KIEE.2026.75.7.1459}
}

@article{xie2026procieee,
  author  = {L. Xie and D. Kim and Q. Zhang and G. Ruan and P. L. Joskow},
  title   = {Large-Scale Data Center Integration in Power Systems: Uncertainty and Multi-Scale Variation},
  journal = {Proc. IEEE},
  note    = {Accepted; preprint arXiv:2610.14264},
  year    = {2026}
}

@misc{wang2026solvrt,
  author       = {W. Wang and A. Bhatnagar and W. Cui and T. Miao and I. Incer},
  title        = {Synthesizing Voltage Ride-Through Controllers for Data Centers},
  howpublished = {arXiv:2608.07289},
  year         = {2026}
}

@misc{norouzi2026dualpq,
  author       = {A. Norouzi and M. Morel},
  title        = {Voltage Ride-Through in Large Loads -- A Dual {PQ} Approach},
  howpublished = {arXiv:2605.00769},
  year         = {2026}
}

@misc{elliott2026sandia,
  author       = {R. Elliott and M. Reno},
  title        = {Voltage and Frequency Ride-Through for Large Computational Loads: State of the Art},
  howpublished = {Sandia National Laboratories presentation (posted by ERCOT)},
  month        = sep,
  year         = {2026}
}

@misc{epri2026rt,
  author       = {{EPRI}},
  title        = {Large Load Ride-Through for Data Centers},
  howpublished = {EPRI Journal},
  month        = aug,
  year         = {2026},
  url          = {https://eprijournal.com/large-load-ride-through-for-data-centers/}
}

@article{gyang2026ojies,
  author  = {P. P. Gyang and P. Chakraborty and L. Meegahapola and X. Yu},
  title   = {Grid-Integrated Data Centers: Architectures, Modeling, Interconnection Requirements, Challenges, and Opportunities},
  journal = {IEEE Open J. Ind. Electron. Soc.},
  year    = {2026},
  doi     = {10.1109/OJIES.2026.3706771}
}

@inproceedings{viththarachchige2026,
  author    = {S. Viththarachchige and D. Alexander and S. Rajendran and A. Manoharan and V. Aravinthan},
  title     = {Concerns and Redefining Load Modeling of Large Data Centers in Future Power Systems---A Review},
  booktitle = {Proc. IEEE Green Technologies Conf. (GreenTech)},
  year      = {2026},
  doi       = {10.1109/GreenTech68285.2026.11471644}
}

@article{gilani2026energies,
  author  = {S. M. W. D. Gilani and W. Nazir and S. J. Rind and M. Jamil},
  title   = {Sustainable Integration of Data Centers into Smart Grids: A Systematic Study of Technologies, Strategies, and Evidence},
  journal = {Energies},
  volume  = {19},
  number  = {22},
  pages   = {5576},
  year    = {2026}
}

@misc{dhakal2026review,
  author       = {B. Dhakal and M. Chamana and B. Gayen and J. Kumar and S. M. S. Rahman and A. Peerzada and K. Y. Lee and O. M. Adeyanju and B. Mitra and S. Kundu and S. Bayne},
  title        = {A Critical Review of Grid Integration Challenges and Flexibility Opportunities for Large-Scale Data Centers},
  howpublished = {arXiv:2610.19374},
  year         = {2026}
}

@article{shrestha2026esm,
  author  = {N. Shrestha and A. Ali and I. Husain and S. Lukic and H. Cui},
  title   = {From Load to Asset: Rethinking {AI} Data Centers for Sustainable Power Grids},
  journal = {IEEE Energy Sustain. Mag.},
  year    = {2026},
  doi     = {10.1109/ESM.2026.3726869}
}

@article{shihab2027apen,
  author  = {M. H. Shihab and others},
  title   = {Grid-Interactive Data Centers for Sustainable and Resilient Energy Systems: Evolution, Trends, and Future Directions},
  journal = {Appl. Energy},
  volume  = {427},
  pages   = {128842},
  year    = {2027},
  doi     = {10.1016/j.apenergy.2026.128842}
}

@article{ghazavi2026pem,
  author  = {M. {Ghazavi Dozein} and P. Mitra and P. Astero and A. Yazdanpanah},
  title   = {Modeling Large Inverter-Based Loads for Power System Dynamic Studies: A Synthesis of Current Practices and Recommendations from the {ESIG} Large Loads Task Force},
  journal = {IEEE Power Energy Mag.},
  year    = {2026},
  doi     = {10.1109/MPE.2026.3722755}
}

@article{chen2026nexus,
  author  = {X. Chen and X. Wang and A. Colacelli and M. Lee and L. Xie},
  title   = {Electricity Demand and Grid Impacts of {AI} Data Centers: Challenges and Prospects},
  journal = {Nexus},
  volume  = {3},
  pages   = {100162},
  year    = {2026},
  doi     = {10.1016/j.ynexs.2026.100162}
}

@inproceedings{seshmasetti2025isgt,
  author    = {V. R. Seshmasetti and A. Ismail and others},
  title     = {A Review of Power System Challenges Stemming from Large {AI} Data Center Loads},
  booktitle = {Proc. IEEE PES ISGT Middle East},
  year      = {2025},
  doi       = {10.1109/ISGTMIDDLEEAST65737.2025.11314244}
}

@article{song2026ojpel,
  author  = {Y. Song and R. Kong and T. Umihara and P. Davari and F. Blaabjerg and others},
  title   = {From Grid to Chip: Power Architecture, Stability, and Flexibility of {AI} Data Centers},
  journal = {IEEE Open J. Power Electron.},
  year    = {2026},
  doi     = {10.1109/OJPEL.2026.3753909}
}

@article{ginzburg2026energies,
  author  = {E. Ginzburg-Ganz and P. Lifshits and R. Machlev and J. Belikov and Z. Krieger and Y. Levron},
  title   = {Technical Challenges of {AI} Data Center Integration into Power Grids---A Survey},
  journal = {Energies},
  volume  = {19},
  number  = {1},
  pages   = {137},
  year    = {2026},
  doi     = {10.3390/en19010137}
}

@misc{mughees2026review,
  author       = {M. Mughees and Y. Li and Y. Chen and Y. R. Li},
  title        = {{AI} Data Centers Meet Electrical Grids: A Review of Power Challenges and Coordinated Solutions},
  howpublished = {arXiv:2610.09060},
  year         = {2026}
}

@article{mohammadi2026adapen,
  author  = {S. Mohammadi and others},
  title   = {Grid Integration of {AI} Data Centers: A Critical Review of Energy Storage Solutions},
  journal = {Adv. Appl. Energy},
  volume  = {23},
  pages   = {100291},
  year    = {2026},
  doi     = {10.1016/j.adapen.2026.100291}
}

@article{s1_surrogate,
  author  = {M. Shamseldein},
  title   = {Solver-Consistent Neural Surrogate for {EMT} Grade {LVRT} Margin Screening of Inverter Dominated Large-Load Interconnections},
  journal = {J. Mod. Power Syst. Clean Energy},
  year    = {2026},
  doi     = {10.35833/MPCE.2026.000612}
}

@techreport{eirgrid2026mpid345,
  author      = {{EirGrid}},
  title       = {Grid Code Modification Recommendation: {MPID}~345 Fault Ride Through, Rate of Change of Frequency and Post Fault Active Power Recovery for Demand Facilities},
  institution = {EirGrid plc},
  address     = {Dublin, Ireland},
  month       = apr,
  year        = {2026},
  url         = {https://cms.eirgrid.ie/sites/default/files/publications/MPID345-Demand-Facilities-FRT-Recommendation-Paper-01042026.pdf}
}

@techreport{eirgrid2026framework,
  author      = {{EirGrid}},
  title       = {Fault Ride Through / Active Power Recovery / Rate of Change of Frequency Grid Code Modification ({MPID}345) Compliance and Derogation Framework Proposal},
  institution = {EirGrid plc},
  address     = {Dublin, Ireland},
  month       = apr,
  year        = {2026},
  url         = {https://cms.eirgrid.ie/sites/default/files/publications/MPID345-Grid-Code-Modification-Compliance-and-Derogation-Process-Framework.pdf}
}

@techreport{soni2026,
  author      = {{SONI}},
  title       = {Consultation Paper: Large Demand Customers Grid Code Modification},
  institution = {System Operator for Northern Ireland},
  month       = jan,
  year        = {2026}
}

@techreport{entsoe2025cnc,
  author      = {{ENTSO-E System Development Committee}},
  title       = {Position on the Need for National Connection Requirements to Ensure {EU} Power System Stability ({CNC}~2.0), Annex A: Connection Requirements for Data Centres},
  institution = {ENTSO-E},
  address     = {Brussels},
  month       = dec,
  year        = {2025}
}

@techreport{acer2025dc,
  author      = {{ACER}},
  title       = {Recommendation No 01/2025, Annex 4: Amended Network Code on Demand Connection},
  institution = {European Union Agency for the Cooperation of Energy Regulators},
  year        = {2025}
}

@techreport{aemc2026final,
  author      = {{Australian Energy Market Commission}},
  title       = {Rule Determination: National Electricity Amendment (Improving the {NEM} Access Standards -- Package 2) Rule 2026},
  institution = {AEMC},
  number      = {ERC0394},
  address     = {Sydney},
  month       = nov,
  year        = {2026},
  url         = {https://www.aemc.gov.au/rule-changes/improving-nem-access-standards-package-2}
}

@techreport{energinet2024tr343,
  author      = {{Energinet}},
  title       = {Technical Regulation 3.4.3: Requirements for Transmission-Connected Demand Facilities, Revision 1},
  institution = {Energinet},
  address     = {Fredericia, Denmark},
  note        = {Effective 1 November 2024},
  year        = {2024}
}

@techreport{ieso2026lcl,
  author      = {{Independent Electricity System Operator}},
  title       = {Technical Requirements for Large Computational Loads Connecting to the {Ontario} Power System, Version 1.0},
  institution = {IESO},
  address     = {Toronto, Canada},
  month       = may,
  year        = {2026},
  url = {https://ieso.ca/-/media/Files/IESO/Document-Library/engage/large-computational-loads/LargeComputationalLoads-20260501-Technical-Requirements.pdf}
}

@techreport{soco2026tcll,
  author      = {{Southern Company Transmission}},
  title       = {Technical Requirements for Transmission-Connected Large Loads ({TCLL}s), v1.4},
  institution = {Southern Company},
  month       = may,
  year        = {2026}
}

@techreport{soco2025ibrmodel,
  author      = {{Southern Company Transmission}},
  title       = {Model Submittal Requirements for Transmission Connected Inverter-Based Resources},
  institution = {Southern Company},
  month       = apr,
  year        = {2025}
}

@misc{neso2025gcdf,
  author       = {R. Rabbani},
  title        = {Large Demand Technical Requirements},
  howpublished = {NESO Grid Code Development Forum, 5 Nov.},
  year         = {2025}
}

@techreport{esig2026perf,
  author      = {{ESIG Large Loads Task Force}},
  title       = {Large Load Performance Requirements: Current Practices and Recommendations},
  institution = {Energy Systems Integration Group},
  address     = {Reston, VA},
  month       = feb,
  year        = {2026},
  url         = {https://www.esig.energy/wp-content/uploads/2026/03/ESIG-Large-Loads-Performance-Requirements-report-2026c.pdf}
}

@techreport{esig2026behav,
  author      = {{ESIG Large Loads Task Force}},
  title       = {Large Loads: Behaviors, Capabilities, and Limitations},
  institution = {Energy Systems Integration Group},
  address     = {Reston, VA},
  month       = jul,
  year        = {2026}
}

@article{cigre2026cse41,
  author  = {J. Matevosyan and B. Badrzadeh and E. Meier and T. K{\"e}r{\c{c}}i and A. Illiceto and J. Deane and Y. Sun and L. Robinson and A. Brochard},
  title   = {Integration of Large Power Electronics-Interfaced Loads in Power Systems: Challenges and Opportunities -- Part {I}: Technical Foundations},
  journal = {CIGRE Science \& Engineering},
  number  = {41},
  month   = jun,
  year    = {2026}
}

@article{arif2018,
  author  = {A. Arif and Z. Wang and J. Wang and B. Mather and H. Bashualdo and D. Zhao},
  title   = {Load Modeling---A Review},
  journal = {IEEE Trans. Smart Grid},
  volume  = {9},
  number  = {6},
  pages   = {5986--5999},
  year    = {2018},
  doi     = {10.1109/TSG.2017.2700436}
}

@article{milanovic2013,
  author  = {J. V. Milanovi{\'c} and K. Yamashita and S. Mart{\'i}nez Villanueva and S. {\v{Z}}. Djokic and L. M. Korunovi{\'c}},
  title   = {International Industry Practice on Power System Load Modeling},
  journal = {IEEE Trans. Power Syst.},
  volume  = {28},
  number  = {3},
  pages   = {3038--3046},
  year    = {2013},
  doi     = {10.1109/TPWRS.2012.2231969}
}

@inproceedings{kosterev2008,
  author    = {D. Kosterev and A. Meklin and J. Undrill and B. Lesieutre and W. Price and D. Chassin and R. Bravo and S. Yang},
  title     = {Load Modeling in Power System Studies: {WECC} Progress Update},
  booktitle = {Proc. IEEE PES General Meeting},
  address   = {Pittsburgh, PA},
  year      = {2008},
  doi = {10.1109/PES.2008.4596557}
}

@techreport{nerc2016dlm,
  author      = {{NERC}},
  title       = {Technical Reference Document: Dynamic Load Modeling},
  institution = {North American Electric Reliability Corporation},
  month       = dec,
  year        = {2016}
}

@inproceedings{sleiman2025cmld,
  author    = {M. Sleiman and A. Sahami and O. Adeosun and N. Rice and K. Vance},
  title     = {Data Center Composite Load Model Parameters' Tuning},
  booktitle = {Proc. CIGRE US National Committee Grid of the Future Symposium},
  address   = {Denver, CO},
  month     = nov,
  year      = {2025}
}

@misc{ali2026steady,
  author       = {M. H. Ali and P. Sang and H. Woo and H. Kang and S. Choi and A. Pandey},
  title        = {Steady-State Equivalent Circuit Model for Data Center Loads},
  howpublished = {arXiv:2608.17925},
  year         = {2026}
}

@misc{syed2026esig,
  author       = {K. Syed},
  title        = {Developer Perspective: Load Modeling and Validation Challenges},
  howpublished = {ESIG Summer Workshop, 16 Jun.},
  year         = {2026}
}

@techreport{nasa7009,
  author      = {{NASA}},
  title       = {Standard for Models and Simulations},
  institution = {National Aeronautics and Space Administration},
  number      = {NASA-STD-7009A},
  year        = {2016}
}

@misc{meier2026nercclo,
  author       = {E. Meier},
  title        = {{NERC} Computational Load Update ({Project} 2026-02)},
  howpublished = {Presentation to ERCOT Large Load Working Group, 21 Aug.},
  year         = {2026}
}

@misc{appa2026nerc,
  author       = {{American Public Power Association}},
  title        = {{NERC} Board Adopts Foundational Computational Load Reliability Standards},
  howpublished = {Public Power Current, Oct.},
  year         = {2026}
}

@misc{ko2026wide,
  author       = {M.-S. Ko and H. Zhu},
  title        = {Wide-Area Power System Oscillations from Large-Scale {AI} Workloads},
  howpublished = {arXiv:2508.16457},
  year         = {2026}
}

@misc{chaudhary2026semimarkov,
  author       = {C. Chaudhary and A. Bera and C. Newlun and M. Benidris and J. Mitra},
  title        = {A Hierarchical Semi-{Markov} Load Model for {AI} Data Centers Coupling Job Scheduling with Bulk-Synchronous-Parallel Power Dynamics},
  howpublished = {arXiv:2607.12222},
  year         = {2026}
}

@inproceedings{sun2019pfc,
  author    = {J. Sun and M. Xu and M. Cespedes and M. Kauffman},
  title     = {Low-Frequency Input Impedance Modeling of Single-Phase {PFC} Converters for Data Center Power System Stability Studies},
  booktitle = {Proc. IEEE Energy Conversion Congr. Expo. (ECCE)},
  year      = {2019}
}

@misc{ercot2026puct,
  author       = {{Electric Reliability Council of Texas}},
  title        = {{PUCT} Report: {NOGRR}~282, Board Priority -- Large Computational Load Ride-Through Requirements},
  howpublished = {Approved by the Public Utility Commission of Texas 9 Jul. 2026; effective 1 Aug. 2026},
  year         = {2026},
  url          = {https://www.ercot.com/files/docs/2026/07/11/282NOGRR-45-PUCT-Report-070926.docx}
}

@misc{cru2026mpid345,
  author       = {{Arthur Cox LLP}},
  title        = {Energy Update: {CRU} Decision on New Requirements for Demand Customers -- Fault Ride Through},
  howpublished = {Summary of the CRU decision of 22 Sep. 2026 approving Grid Code modification MPID~345},
  month        = sep,
  year         = {2026},
  url          = {https://www.arthurcox.com/insights/energy-update-cru-decision-on-new-requirements-for-demand-customers-fault-ride-through/}
}

@inproceedings{yogarathnam2026bess,
  author    = {A. Yogarathnam and D. Cheerla and D. Jain and S. Natti},
  title     = {Auxiliary Control Strategies for Co-Located Battery Energy Storage Systems to Enhance {LVRT} and Load Smoothing in Large Data Centers},
  booktitle = {Proc. IEEE Int. Conf. Commun., Control, Comput. Technol. Smart Grids (SmartGridComm)},
  address   = {College Station, TX, USA},
  year      = {2026}
}

@misc{cheerla2026emtrms,
  author       = {D. Cheerla and A. Yogarathnam and D. Jain and S. Natti},
  title        = {Control-Layer Interpretation of Reactive Power Recovery Limits in {EMT}--{RMS} Model Benchmarking of Inverter-Based Resources},
  howpublished = {arXiv:2610.41221; accepted to IEEE PES Grid Edge Technologies Conf. 2027},
  year         = {2026}
}

@misc{cheerla2026lvrt,
  author       = {D. Cheerla and A. Yogarathnam and D. Jain and S. Natti},
  title        = {{EMT} Assessment of {SCR} and {X/R} Effects on Current-Priority-Dependent {LVRT} Response of a Grid-Following {BESS}},
  howpublished = {arXiv:2610.41212; accepted to IEEE Texas Power and Energy Conf. (TPEC) 2027},
  year         = {2026}
}

@inproceedings{lu2026,
  author = {S. Lu and C. Xiao and Y. Weng},
  title = {{A Unified Dynamic Modeling and Data-Driven Workload Framework for Large Electric Loads}},
  booktitle = {Proc. IEEE Power \& Energy Society General Meeting (PESGM)},
  year = {2026},
  pages = {1--5},
  doi = {10.1109/pesgm58988.2026.11693764}
}

@inproceedings{maheshwari2026,
  author = {L. Maheshwari and H. Dai and B. Sarlioglu and R. Chakraborty and Y. Xia and M. Harper},
  title = {{AI Data Centers: A Review of Power Quality Challenges and Mitigation Strategies}},
  booktitle = {Proc. IEEE Power and Energy Conf. at Illinois (PECI)},
  year = {2026},
  pages = {1--6},
  doi = {10.1109/peci70026.2026.11516457}
}

@inproceedings{jimenezortega2026,
  author = {E. Jiménez-Ortega and S. Santoso and R. F. Arritt},
  title = {{Analysis of Flicker Caused by Large Electronic Loads in Weak Power Grids}},
  booktitle = {Proc. IEEE Power \& Energy Society General Meeting (PESGM)},
  year = {2026},
  pages = {1--5},
  doi = {10.1109/pesgm58988.2026.11693691}
}

@inproceedings{azizi2026,
  author = {A. Azizi and A. Babaei and A. Aluko and R. Salehi and H. Bayat},
  title = {{Analysis of Sub-Synchronous Oscillations Induced by AI Data Centers and Damping Enhancement Using Grid-Forming Inverters}},
  booktitle = {Proc. IEEE Power \& Energy Society General Meeting (PESGM)},
  year = {2026},
  pages = {1--5},
  doi = {10.1109/pesgm58988.2026.11693096}
}

@article{shamseldein2026,
  author = {M. Shamseldein},
  title = {{Co-Located AI data centers, hydrogen production, and battery storage: a multi-timescale grid-connection assessment}},
  journal = {Global Energy Interconnection},
  year = {2026},
  doi = {10.1016/j.gloei.2026.08.002}
}

@inproceedings{liu2025,
  author = {Q. Liu and X. Jia and S. Biao and S. Dulal and C. Shen and Y. Liu and others},
  title = {{Detection of Data Center Load Loss Event with Wide-Area-Monitoring-System}},
  booktitle = {Proc. IEEE PES Innovative Smart Grid Technologies Conf. Europe (ISGT Europe)},
  year = {2025},
  pages = {1--5},
  doi = {10.1109/isgteurope64741.2025.11305297}
}

@inproceedings{sun2021,
  author = {J. Sun and S. Wang and J. Wang and L. M. Tolbert},
  title = {{Development of a Converter-Based Data Center Power Emulator}},
  booktitle = {Proc. IEEE Applied Power Electronics Conf. Expo. (APEC)},
  year = {2021},
  pages = {126--133},
  doi = {10.1109/apec42165.2021.9487043}
}

@article{sun2022,
  author = {J. Sun and S. Wang and J. Wang and L. M. Tolbert},
  title = {{Dynamic Model and Converter-Based Emulator of a Data Center Power Distribution System}},
  journal = {IEEE Trans. Power Electron.},
  year = {2022},
  volume = {37},
  number = {7},
  pages = {8420--8432},
  doi = {10.1109/tpel.2022.3146354}
}

@inproceedings{arifin2026,
  author = {M. H. Arifin and S. F. Khan and S. Kamalasadan and M. Smith and Y. P. Agalgaonkar and R. Singh},
  title = {{Dynamic Modeling of Large Data Centers Integrated Electric Grid With Impact Analysis and Oscillatory Behavior Assessment}},
  booktitle = {Proc. IEEE Rural Electric Power Conf. (REPC)},
  year = {2026},
  pages = {1--6},
  doi = {10.1109/repc71040.2026.00010}
}

@inproceedings{reyes2024,
  author = {G. E. C. Reyes and C. Tomlin and R. Dye and D. Callaway},
  title = {{Effects of Dynamic Power Electronic Load Models on Power Systems Analysis Using ZIP-E Loads}},
  booktitle = {Proc. North American Power Symp. (NAPS)},
  year = {2024},
  pages = {1--6},
  doi = {10.1109/naps61145.2024.10741785}
}

@inproceedings{afzaal2025,
  author = {M. U. Afzaal and Y. Liu and Y. Yang and C. Shen and Q. Liu and Y. Wu and others},
  title = {{EMT-Based Analysis of Supraharmonic Emission and Propagation in IBRs and Data Centers Dominated Power Systems}},
  booktitle = {Proc. New Jersey Future Energy Transmission Conf. (NJFET)},
  year = {2025},
  pages = {1--6},
  doi = {10.1109/njfet67489.2025.11380685}
}

@inproceedings{biswas2026,
  author = {S. Biswas and A. C. Varghese and K. Chatterjee and S. Nekkalapu and B. Ross and J. Follum},
  title = {{Evaluating the Risk to Bulk Power System Reliability from Large Load Induced Oscillations}},
  booktitle = {Proc. IEEE PES Transmission and Distribution Conf. Expo. (T\&D)},
  year = {2026},
  pages = {1--5},
  doi = {10.1109/td48022.2026.11562229}
}

@article{thomas2026,
  author = {K. Thomas and J. Gumpert and L. Jia and N. Mohan and K. Syed},
  title = {{Evolving the Large Load Interconnection Process: Challenges, Lessons Learned, and Opportunities From the United States}},
  journal = {IEEE Power Energy Mag.},
  year = {2026},
  volume = {24},
  number = {6},
  pages = {68--83},
  doi = {10.1109/mpe.2026.3713098}
}

@article{hussein2026,
  author = {H. M. Hussein and O. A. Mohammed},
  title = {{Hyperscale Loads and Energy Storage: A Grid Code Compliance Perspective}},
  journal = {Electronics},
  year = {2026},
  volume = {15},
  number = {12},
  pages = {2669},
  doi = {10.3390/electronics15122669}
}

@article{zhu2023,
  author = {T. Zhu and X. Wang and F. Zhao and G. V. Torrico-Bascopé},
  title = {{Impedance-Based Aggregation of Paralleled Power Factor Correction Converters in Data Centers}},
  journal = {IEEE Trans. Power Electron.},
  year = {2023},
  volume = {38},
  number = {4},
  pages = {5254--5265},
  doi = {10.1109/tpel.2022.3230645}
}

@inproceedings{aghadinuno2026,
  author = {C. Aghadinuno and S. Ahmed and M. Alamaniotis and B. Wang and N. Gatsis},
  title = {{Investigation of AI Data Center Load Impact on Power System Frequency Using Real-World Datasets}},
  booktitle = {Proc. IEEE Green Technologies Conf. (GreenTech)},
  year = {2026},
  pages = {1--6},
  doi = {10.1109/greentech68285.2026.11471548}
}

@article{rashwan2026,
  author = {A. Rashwan and B. Guha and M. Shamseldein and S. Kynev and T. Meridji and A. Dadashzade and others},
  title = {{Large Electronic Loads: Understanding and Mitigating Risks to the Power System}},
  journal = {IEEE Power Energy Mag.},
  year = {2026},
  volume = {24},
  number = {6},
  pages = {84--95},
  doi = {10.1109/mpe.2026.3716964}
}

@inproceedings{ahmad2025,
  author = {G. Ahmad and M. Shafiullah and N. M. Al-Yazidi and M. Zaery and M. Al-Dhaifallah and M. A. Abido and others},
  title = {{Load Modeling for Cryptocurrency Mining Devices Using System Identification and Machine Learning}},
  booktitle = {Proc. IEEE Int. Multi-Conf. on Systems, Signals and Devices},
  year = {2025},
  pages = {516--521},
  doi = {10.1109/ssd64182.2025.10989872}
}

@inproceedings{zubi2024,
  author = {H. Zubi and N. Hfouda},
  title = {{Measurement-Based Data Center Complete Load Model Construction \& Power Quality Assessment at Distribution Level}},
  booktitle = {Proc. IEEE Int. Maghreb Meeting Conf. Sci. Tech. Autom. Control Comput. Eng. (MI-STA)},
  year = {2024},
  pages = {333--340},
  doi = {10.1109/mi-sta61267.2024.10599677}
}

@inproceedings{chaudhary2026,
  author = {C. Chaudhary and M. Murillo and M. Ben-Idris and J. Mitra and D. Pandit and A. Bera},
  title = {{Modal Analysis of Spatial Load Correlation in AI Data Center-Dominated Power Systems}},
  booktitle = {Proc. Int. Conf. on Smart Energy Systems and Technologies (SEST)},
  year = {2026},
  pages = {1--6},
  doi = {10.1109/sest67798.2026.11712308}
}

@inproceedings{tang2026,
  author = {J. Tang and B. Xie and X. Xu and G. Repalle and B. Choi and L. Huang and others},
  title = {{Modeling and Simulation of Data Center Loads in PJM System Stability Assessment}},
  booktitle = {Proc. IEEE Power \& Energy Society General Meeting (PESGM)},
  year = {2026},
  pages = {1--5},
  doi = {10.1109/pesgm58988.2026.11693491}
}

@article{ramasubramanian2025,
  author = {D. Ramasubramanian and P. Mitra and R. O’Keefe and J. Fonseka and B. Jayasekara and M. Dewadasa and others},
  title = {{Modeling of Load for Transmission Systems Analysis: Aggregate versus detailed and positive sequence vs electromagnetic transient [Feature]}},
  journal = {IEEE Power Energy Mag.},
  year = {2025},
  volume = {23},
  number = {4},
  pages = {81--98},
  doi = {10.1109/mpe.2024.3511454}
}

@inproceedings{sreejith2026,
  author = {A. Sreejith and B. Natarajan},
  title = {{OALS: An Open source AI-Datacenter Load Simulator for Grid Impact Analysis}},
  booktitle = {Proc. IEEE Power \& Energy Society General Meeting (PESGM)},
  year = {2026},
  pages = {1--5},
  doi = {10.1109/pesgm58988.2026.11693512}
}

@inproceedings{sanati2026,
  author = {S. Sanati and A. Mahari and M. Martynov and P. Parikh},
  title = {{Power Swing-Induced Maloperation in Distance Protection with Large Data Center Load}},
  booktitle = {Proc. IEEE Power \& Energy Society General Meeting (PESGM)},
  year = {2026},
  pages = {1--5},
  doi = {10.1109/pesgm58988.2026.11693548}
}

@inproceedings{samanta2024,
  author = {A. Samanta and Q. Zhang and L. Xie},
  title = {{Prediction of Cryptocurrency Mining Load Tripping Through Learning-Based Fault Classification}},
  booktitle = {Proc. IEEE Texas Power and Energy Conf.},
  year = {2024},
  doi = {10.1109/tpec60005.2024.10472233}
}

@inproceedings{meier2026,
  author = {E. Meier and S. Basumallik},
  title = {{Reliability Challenges and Solutions for Large Load Integration in Bulk Power Systems}},
  booktitle = {Proc. IEEE PES Transmission and Distribution Conf. Expo. (T\&D)},
  year = {2026},
  pages = {1--5},
  doi = {10.1109/td48022.2026.11562016}
}

@article{chaudhary2026b,
  author = {C. Chaudhary and A. Abdelkader and M. Ben-Idris and J. Mitra},
  title = {{Spatially Correlated Load Dynamics in AI Data-Center-Dominated Grids}},
  journal = {IEEE Trans. Power Syst.},
  year = {2026},
  volume = {PP},
  number = {99},
  pages = {1--18},
  doi = {10.1109/tpwrs.2026.3729445}
}

@inproceedings{shen2025,
  author = {W. Shen and M. Zhu and T. Ding and M. Gao and Y. Jiao},
  title = {{Sub-Synchronous Oscillation Characteristics Analysis in Distribution Networks with UPS Clusters Based on Impedance Modeling}},
  booktitle = {Proc. Int. Conf. on Power and Energy Technology (ICPET)},
  year = {2025},
  pages = {55--60},
  doi = {10.1109/icpet66029.2025.11160368}
}

@inproceedings{barati2026,
  author = {M. Barati},
  title = {{Subsynchronous Oscillation Assessment in Grid Integration of Large Data Center Loads}},
  booktitle = {Proc. IEEE Power \& Energy Society General Meeting (PESGM)},
  year = {2026},
  pages = {1--5},
  doi = {10.1109/pesgm58988.2026.11693575}
}

@article{conto2025,
  author = {J. Conto and Y. Cheng and J. Rose and J. Schmall},
  title = {{Texas Loads Ride Toward Grid Stability: Voltage Ride Through of Large Power Electronic Loads}},
  journal = {IEEE Power Energy Mag.},
  year = {2025},
  volume = {23},
  number = {5},
  pages = {56--67},
  doi = {10.1109/mpe.2025.3564963}
}

@inproceedings{zechevikj2026,
  author = {D. Zechevikj and S. Santoso},
  title = {{Understanding the Impact of Autoreclosing on Voltage Ride-Through of Large Electronic Loads}},
  booktitle = {Proc. IEEE Power \& Energy Society General Meeting (PESGM)},
  year = {2026},
  pages = {1--5},
  doi = {10.1109/pesgm58988.2026.11693652}
}

@inproceedings{khomarudin2025,
  author = {R. Khomarudin and R. A. A. Hasibi},
  title = {{Voltage Dips Mitigation due to Under Voltage Relay Trip in Data Center}},
  booktitle = {Proc. Int. Conf. on Electronic and Electrical Engineering and Intelligent System (ICE3IS)},
  year = {2025},
  pages = {333--338},
  doi = {10.1109/ice3is66769.2025.11280939}
}

@article{miah2026,
  author = {M. S. Miah and M. U. Aslam and R. Shah and N. Amjady and M. R. Islam and N. Mohammed},
  title = {{Voltage Stability Assessment and Enhancement in Power Systems With Large-Scale AI Data Centre Loads}},
  journal = {IET Gener. Transm. Distrib.},
  year = {2026},
  volume = {20},
  number = {1},
  pages = {e70469},
  doi = {10.1049/gtd2.70469}
}

@inproceedings{gu2026,
  author = {L. Gu and A. Nabih and S. Kudva and M. Mosa and R. Abramson and Y. Elasser and others},
  title = {{Active Power Buffer for Smoothing GPU Workload Transients in Data Center Power Delivery}},
  booktitle = {Proc. IEEE Applied Power Electronics Conf. Expo. (APEC)},
  year = {2026},
  pages = {158--163},
  doi = {10.1109/apec51134.2026.11516732}
}

@inproceedings{zhabelova2015,
  author = {G. Zhabelova and A. Yavarian and V. Vyatkin},
  title = {{Data center power dynamics within the settings of regional power grid}},
  booktitle = {Proc. IEEE 20th Conf. on Emerging Technologies \& Factory Automation (ETFA)},
  year = {2015},
  pages = {1--5},
  doi = {10.1109/etfa.2015.7301633}
}

@inproceedings{sun2025,
  author = {R. Sun and Q. Liu and J. Zhang and D. Liu and H. Yue and G. Wang and others},
  title = {{Data-Driven Inverse Optimization for Frequency Response Capability Modeling of Data Center}},
  booktitle = {Proc. IEEE Int. Conf. on Energy Technologies for Future Grids},
  year = {2025},
  doi = {10.1109/etfg61999.2025.11401249}
}

@article{kim2025,
  author = {S. Kim and S. Jung and J. Lee and S. Oh and J. Shin},
  title = {{Development of Fast Response Technology Using Data Center On-Line UPS}},
  journal = {Trans. Korean Inst. Electr. Eng.},
  year = {2025},
  volume = {74},
  number = {11},
  pages = {1817--1823},
  doi = {10.5370/kiee.2025.74.11.1817}
}

@inproceedings{badakhshan2026,
  author = {S. Badakhshan and R. A. Jacob and A. M. Rad and C. Pan and Y. Li and J. Zhang},
  title = {{Dynamic Stability Assessment of Grid-Connected Data Centers Powered by Small Modular Reactors}},
  booktitle = {Proc. IEEE Power \& Energy Society General Meeting (PESGM)},
  year = {2026},
  pages = {1--5},
  doi = {10.1109/pesgm58988.2026.11693234}
}

@article{wang2026,
  author = {Z. Wang and Y. Zou and J. Wang and H. Chen and J. Qian and C. Gong},
  title = {{Electricity computing collaboration: Data center power configuration, load modeling and prediction, power dispatching}},
  journal = {J. Electron. Meas. Instrum.},
  year = {2026},
  volume = {40},
  number = {5},
  pages = {1--14},
  doi = {10.13382/j.jemi.b2508945}
}

@inproceedings{ghosh2025,
  author = {S. Ghosh and M. A. H. Sadi},
  title = {{Enhancing Grid Resilience for Giga-Watt Scale Data Centers Using High Voltage Circuit Breaker Operated Braking Resistors}},
  booktitle = {Proc. IEEE Int. Conf. on Energy Technologies for Future Grids (ETFG)},
  year = {2025},
  pages = {1--6},
  doi = {10.1109/etfg61999.2025.11401157}
}

@inproceedings{harzevili2026,
  author = {M. N. Harzevili and R. Hadidi and N. S. Ghiasi and Z. H. Smith},
  title = {{Evolution of Load Modeling in Power Systems: From Classical Representations to Data-Driven and Physics-Informed Models}},
  booktitle = {Proc. IEEE/IAS 62nd Industrial and Commercial Power Systems Technical Conf. (I\&CPS)},
  year = {2026},
  pages = {1--9},
  doi = {10.1109/icps67768.2026.11712982}
}

@article{tao2025,
  author = {X. Tao and R. Gadh},
  title = {{Fast Frequency Response Potential of Data Centers Through Workload Modulation and UPS Coordination}},
  journal = {IEEE Access},
  year = {2025},
  volume = {13},
  pages = {214511--214519},
  doi = {10.1109/access.2025.3646120}
}

@inproceedings{azizi2025,
  author = {A. Azizi and A. Aluko and S. Maleki and R. Salehi and H. Bayat},
  title = {{Grid-Forming Inverter Applications for Improved Reliability and Power Quality in AI Data Centers}},
  booktitle = {Proc. IEEE Power \& Energy Society General Meeting (PESGM)},
  year = {2025},
  pages = {1--5},
  doi = {10.1109/pesgm52009.2025.11225523}
}

@article{shamseldein2027,
  author = {M. Shamseldein},
  title = {{Impedance-passivity coordination of hydrogen-coupled AI data centers for oscillation damping and fault ride-through at transmission interfaces}},
  journal = {Electr. Power Syst. Res.},
  year = {2027},
  volume = {262},
  pages = {113557},
  doi = {10.1016/j.epsr.2026.113557}
}

@inproceedings{kundu2026,
  author = {S. Kundu and K. Chatterjee and R. R. Hossain and S. P. Nandanoori and V. Adetola},
  title = {{Managing Risks from Large Digital Loads Using Coordinated Grid-Forming Storage Network}},
  booktitle = {Proc. IEEE PES Transmission and Distribution Conf. Expo. (T\&D)},
  year = {2026},
  pages = {1--5},
  doi = {10.1109/td48022.2026.11562446}
}

@article{shamseldein2026b,
  author = {M. Shamseldein},
  title = {{Mitigating Grid Stability Risks From Large AI Data Centres: A Three-Mode Supervisory Control Framework for Centralized UPS Systems}},
  journal = {IET Gener. Transm. Distrib.},
  year = {2026},
  volume = {20},
  number = {1},
  pages = {e70369},
  doi = {10.1049/gtd2.70369}
}

@inproceedings{lyu2026,
  author = {X. Lyu and W. Du and S. M. Mohiuddin and B. A. Ross},
  title = {{Mitigating Rapid AI Data Center Power Fluctuations with Grid-Forming Battery Storage: Accounting for Communication Delays}},
  booktitle = {Proc. IEEE Power \& Energy Society General Meeting (PESGM)},
  year = {2026},
  pages = {1--5},
  doi = {10.1109/pesgm58988.2026.11693692}
}

@inproceedings{silva2026,
  author = {G. E. d. Silva and R. Torquato and W. Freitas},
  title = {{Probabilistic Power Demand Model For Colocation Data Centers}},
  booktitle = {Proc. IEEE Power \& Energy Society General Meeting (PESGM)},
  year = {2026},
  pages = {1--5},
  doi = {10.1109/pesgm58988.2026.11693517}
}

@inproceedings{barati2025,
  author = {M. Barati},
  title = {{Resilience to Dynamic Load Attacks Under AI Demand and Hyperscale Data Centers}},
  booktitle = {Proc. IEEE 7th Int. Conf. on Trust, Privacy and Security in Intelligent Systems, and Applications},
  year = {2025},
  pages = {445--454},
  doi = {10.1109/tps-isa67132.2025.00054}
}

@inproceedings{ahmed2021,
  author = {K. M. U. Ahmed and R. A. d. Oliveira and M. H. J. Bollen and M. Alvarez},
  title = {{Risk Assessment of Server Outages Due to Voltage Dips in the Internal Power Supply System of a Data Center}},
  booktitle = {Proc. 26th Int. Conf. Electricity Distribution (CIRED)},
  year = {2021},
  pages = {708--712},
  doi = {10.1049/icp.2021.1846}
}

@inproceedings{parikh2025,
  author = {P. Parikh and M. Ali and N. Virani and S. C. Evans},
  title = {{Strain to Synergy: Powering AI Data Centers with Sustainable and Smart Power Grid Infrastructures}},
  booktitle = {Proc. IEEE Int. Conf. on Communications, Control, and Computing Technologies for Smart Grids},
  year = {2025},
  doi = {10.1109/smartgridcomm65349.2025.11204571}
}

@article{azizi2026b,
  author = {A. Azizi and S. Morovati and A. Zamani and J. Piruzza and D. Guo and Z. Liu and others},
  title = {{Strengthening Data Center Operations Using Grid-Forming Battery Energy Storage as a Line-Interactive Uninterruptible Power Supply}},
  journal = {Int. J. Electr. Power Energy Syst.},
  year = {2026},
  volume = {175},
  pages = {111638},
  doi = {10.1016/j.ijepes.2026.111638}
}

@inproceedings{seo2026,
  author = {S. W. Seo and S. Pannala and B. Poudel and R. Hovsapian},
  title = {{Supercapacitor Assisted Power Conditioning System to Enhance Ride-through Functionality of Datacenter Loads}},
  booktitle = {Proc. IEEE PES Transmission and Distribution Conf. Expo. (T\&D)},
  year = {2026},
  pages = {1--5},
  doi = {10.1109/td48022.2026.11562372}
}

@inproceedings{hossain2025,
  author = {M. S. Hossain and M. M. Rahman and M. R. Islam and D. Sutanto and K. M. Muttaqi},
  title = {{The Impact of Large-Scale Data Center Load Integration into Electrical Power Networks}},
  booktitle = {Proc. IEEE Int. Conf. on Energy Technologies for Future Grids (ETFG)},
  year = {2025},
  pages = {1--6},
  doi = {10.1109/etfg61999.2025.11401166}
}

@inproceedings{ahmed2022,
  author = {K. M. U. Ahmed and M. H. J. Bollen and M. Alvarez and S. S. Letha},
  title = {{The Impacts of Voltage Disturbances Due to Faults In the Power Supply System of A Data Center}},
  booktitle = {Proc. Int. Conf. on Harmonics \& Quality of Power (ICHQP)},
  year = {2022},
  pages = {1--6},
  doi = {10.1109/ichqp53011.2022.9808804}
}

@inproceedings{ilic2026,
  author = {M. Ilić},
  title = {{The Key Role of FACTS Control in Enabling Integration of Large Data Centers}},
  booktitle = {Proc. IEEE PES Transmission and Distribution Conf. Expo. (T\&D)},
  year = {2026},
  pages = {1--5},
  doi = {10.1109/td48022.2026.11561924}
}

@article{ross2026,
  author = {B. A. Ross and X. Lyu and U. C. Nwaneto and S. M. Mohiuddin and A. B. Nassif},
  title = {{Using Grid-Forming Energy Storage Systems to Provide Dynamic Active Power Support for Hyperscale Data Center}},
  journal = {IEEE Access},
  year = {2026},
  volume = {14},
  pages = {29250--29259},
  doi = {10.1109/access.2026.3666746}
}

@article{dorji2026pdkan,
  author = {S. Dorji and Y. Zhang and G. Nourbakhsh and Y. Mishra},
  title = {{Physics-informed data-driven Kolmogorov–Arnold networks for dynamic load modelling in power systems}},
  journal = {Int. J. Electr. Power Energy Syst.},
  volume = {182},
  pages = {112207},
  year = {2026},
  doi = {10.1016/j.ijepes.2026.112207}
}

@article{lin2026rlsr,
  author = {D. Lin and H. Guo and J. Wang and M. Yue and T. Zhao},
  title = {{Reinforcement Learning-Based Symbolic Regression for Load Modeling}},
  journal = {IEEE Open Access J. Power Energy},
  volume = {13},
  pages = {250--261},
  year = {2026},
  doi = {10.1109/oajpe.2026.3672033}
}

@article{qiu2026bayes,
  author = {Y. Qiu and Y. Xu and W. Gu and X. Ye and L. Mili and Y. Yao and Q. Wang and S. Lu},
  title = {{Decentralized Analytic Bayesian Probabilistic Composite Load Modeling: Sensitivity, Identifiability and Interpretability}},
  journal = {IEEE Trans. Power Syst.},
  volume = {41},
  number = {6},
  pages = {4791--4805},
  year = {2026},
  doi = {10.1109/tpwrs.2026.3724316}
}

@article{acero2026processes,
  author = {J. C. R. Acero and R. Isaza-Ruget and J. Rosero-García},
  title = {{Reduced-Order Modeling of Transient Events in Data Centers Using Dimensionality Reduction Techniques}},
  journal = {Processes},
  volume = {14},
  number = {10},
  pages = {1665},
  year = {2026},
  doi = {10.3390/pr14101665}
}

@article{transtrum2025tailoring,
  author = {M. K. Transtrum and A. M. Stankovic and C. C. Youn},
  title = {{Tailoring and verification of dynamic load models in power system studies: the WECC load example}},
  journal = {Sustain. Energy Grids Netw.},
  volume = {44},
  pages = {101970},
  year = {2025},
  doi = {10.1016/j.segan.2025.101970}
}

@article{cicilio2020metrics,
  author = {P. Cicilio and E. Cotilla-Sanchez},
  title = {{Evaluating Measurement-Based Dynamic Load Modeling Techniques and Metrics}},
  journal = {IEEE Trans. Power Syst.},
  volume = {35},
  number = {3},
  pages = {1805--1811},
  year = {2020},
  doi = {10.1109/tpwrs.2019.2949722}
}

@article{zhu2026diffusion,
  author = {F. Zhu and D. Torbunov and Z. Jiang and T. Zhao and A. Yogarathnam and Y. Ren and M. Yue},
  title = {{Diffusion model-based parameter estimation in dynamic power systems}},
  journal = {Commun. Eng.},
  volume = {5},
  number = {1},
  pages = {123},
  year = {2026},
  doi = {10.1038/s44172-026-00670-z}
}

@inproceedings{qi2022pmu,
  author = {C. Qi and X. Wang},
  title = {{Comparative Study of Data-Driven Dynamic Load Model Identification Methods Based on Simulated and Actual PMU Data}},
  booktitle = {Proc. North American Power Symp. (NAPS)},
  pages = {1--6},
  year = {2022},
  doi = {10.1109/naps56150.2022.10012264}
}

@article{mitrentsis2021adn,
  author = {G. Mitrentsis and H. Lens},
  title = {{Data-Driven Dynamic Models of Active Distribution Networks Using Unsupervised Learning Techniques on Field Measurements}},
  journal = {IEEE Trans. Smart Grid},
  volume = {12},
  number = {4},
  pages = {2952--2965},
  year = {2021},
  doi = {10.1109/tsg.2021.3057763}
}

@inproceedings{lin2024taiwan,
  author = {Y. Lin and S. Lee and C. Chu},
  title = {{Experiences and Lessons Learned from Load Modelling for and Applications to Taiwan Power Systems}},
  booktitle = {Proc. IEEE/IAS Industrial and Commercial Power Systems Technical Conf. (I\&CPS)},
  year = {2024},
  doi = {10.1109/icps60943.2024.10563877}
}

@article{paidi2020ocl,
  author = {E. S. N. R. Paidi and A. Nechifor and M. M. Albu and J. Yu and V. Terzija},
  title = {{Development and Validation of a New Oscillatory Component Load Model for Real-Time Estimation of Dynamic Load Model Parameters}},
  journal = {IEEE Trans. Power Del.},
  volume = {35},
  number = {2},
  pages = {618--629},
  year = {2020},
  doi = {10.1109/tpwrd.2019.2918059}
}

@inproceedings{leinakse2021dfr,
  author = {M. Leinakse and J. Kilter},
  title = {{Processing and Filtering Digital Fault Recorder Events for Load Model Estimation}},
  booktitle = {Proc. IEEE PES Innovative Smart Grid Technol. Europe (ISGT Europe)},
  year = {2021},
  doi = {10.1109/isgteurope52324.2021.9640026}
}

@article{bu2020edrl,
  author = {F. Bu and Z. Ma and Y. Yuan and Z. Wang},
  title = {{WECC Composite Load Model Parameter Identification Using Evolutionary Deep Reinforcement Learning}},
  journal = {IEEE Trans. Smart Grid},
  volume = {11},
  number = {6},
  pages = {5407--5417},
  year = {2020},
  doi = {10.1109/tsg.2020.3008730}
}

@article{wang2020ddqn,
  author = {X. Wang and Y. Wang and D. Shi and J. Wang and Z. Wang},
  title = {{Two-Stage WECC Composite Load Modeling: A Double Deep Q-Learning Networks Approach}},
  journal = {IEEE Trans. Smart Grid},
  volume = {11},
  number = {5},
  pages = {4331--4344},
  year = {2020},
  doi = {10.1109/tsg.2020.2988171}
}

@article{xiao2023neuralode,
  author = {T. Xiao and Y. Chen and S. Huang and T. He and H. Guan},
  title = {{Feasibility Study of Neural ODE and DAE Modules for Power System Dynamic Component Modeling}},
  journal = {IEEE Trans. Power Syst.},
  volume = {38},
  number = {3},
  pages = {2666--2678},
  year = {2023},
  doi = {10.1109/tpwrs.2022.3194570}
}

@article{shen2024neudye,
  author = {Q. Shen and Y. Zhou and Q. Zhang and S. Maslennikov and X. Luo and P. Zhang},
  title = {{Physics-Aware Neural Dynamic Equivalence of Power Systems}},
  journal = {IEEE Trans. Power Syst.},
  volume = {39},
  number = {1},
  pages = {2341--2344},
  year = {2024},
  doi = {10.1109/tpwrs.2023.3328162}
}

@inproceedings{zhang2024lstmdcdc,
  author = {Z. Zhang and G. Wang and X. Liu},
  title = {{A Black-Box Modeling Method Based on LSTM Network for DC-DC Converters}},
  booktitle = {Proc. CPSS \& IEEE Int. Symp. Energy Storage and Conversion (ISESC)},
  pages = {437--442},
  year = {2024},
  doi = {10.1109/isesc63657.2024.10785341}
}

@inproceedings{varshney2025sysid,
  author = {H. Varshney and H. Jain},
  title = {{A System Identification Approach for Developing Dynamic Models of Inverter-Based Motor Loads}},
  booktitle = {Proc. IEEE Int. Conf. Energy Technologies for Future Grids (ETFG)},
  year = {2025},
  doi = {10.1109/etfg61999.2025.11402362}
}

@article{guruwacharya2024phil,
  author = {N. Guruwacharya and S. Chakraborty and G. Saraswat and R. Bryce and T. M. Hansen and R. Tonkoski},
  title = {{Data-Driven Modeling of Grid-Forming Inverter Dynamics Using Power Hardware-in-the-Loop Experimentation}},
  journal = {IEEE Access},
  volume = {12},
  pages = {52267--52281},
  year = {2024},
  doi = {10.1109/access.2024.3383323}
}

@article{subedi2025overview,
  author = {S. Subedi and Y. Gui and Y. Xue},
  title = {{Applications of Data-Driven Dynamic Modeling of Power Converters in Power Systems: An Overview}},
  journal = {IEEE Trans. Ind. Appl.},
  volume = {61},
  number = {2},
  pages = {2434--2456},
  year = {2025},
  doi = {10.1109/tia.2025.3529797}
}

@article{singh2025aireview,
  author = {S. D. Singh and M. D. Singh and A. K. Singh and S. Kumar},
  title = {{AI-Enhanced and Other Load Modelling in Modern Power Systems: A Comprehensive Review of Advances, Challenges, and Future Directions}},
  journal = {Rev. Int. M\'etodos Num\'er. C\'alc. Dise\~no Ing.},
  volume = {41},
  number = {4},
  pages = {77},
  year = {2025},
  doi = {10.23967/j.rimni.2025.10.71136}
}

@article{lu2025uniqueness,
  author = {S. Lu and J. Ding and M. Chen and W. Gu and J. Zhu and Y. Xu and Z. Dong and Z. Sun},
  title = {{On the Solution Uniqueness of Data-Driven Modeling of Flexible Loads}},
  journal = {IEEE Trans. Smart Grid},
  volume = {16},
  number = {2},
  pages = {1993--1996},
  year = {2025},
  doi = {10.1109/tsg.2024.3518094}
}

@inproceedings{avila2020cmld,
  author = {N. F. Avila and L. Callegaro and J. Fletcher},
  title = {{Measurement-based parameter estimation for the WECC composite load model with distributed energy resources}},
  booktitle = {Proc. IEEE PES General Meeting},
  volume = {2020-August},
  pages = {9281843},
  year = {2020},
  doi = {10.1109/pesgm41954.2020.9281843}
}

@inproceedings{govind2023hpc,
  author = {A. Govind and S. Bhalachandra and Z. Zhao and E. Rrapaj and B. Austin and H. A. Nam},
  title = {{Comparing Power Signatures of HPC Workloads: Machine Learning vs Simulation}},
  booktitle = {Proc. SC'23 Workshops (ACM)},
  pages = {1890--1893},
  year = {2023},
  doi = {10.1145/3624062.3624274}
}

@misc{lin2026identifiability,
  author = {D. Lin and J. Wang and M. Yue and T. Zhao},
  title = {{Identifiability-Guided Composite Load Modeling of AI Data Centers}},
  howpublished = {arXiv:2610.19565},
  year = {2026}
}

@misc{saroare2026dcclm,
  author = {M. K. Saroare and M. R. Ahmed and A. Hussain},
  title = {{DC-CLM: Extending the WECC Composite Load Model for AI Data Center Dynamics}},
  howpublished = {arXiv:2609.22093},
  year = {2026}
}

@misc{woo2026reinsertion,
  author = {J. H. Woo and V. D. Paduani and Z. Yang and Y. Ma and Q. Xiao and N. Lu},
  title = {{Reinsertion-Aware Parameter Estimation of the WECC Composite Load Model via Closed-Loop Equivalent-Network Co-Estimation}},
  howpublished = {arXiv:2610.34497},
  year = {2026}
}

@misc{shenoy2026switched,
  author = {V. Shenoy and D. Ajeyemi and M. Abdelgalil and G. Cavraro and E. Dall'Anese and J. I. Poveda},
  title = {{On the Stability of Power Systems with AI Data Centers: A Multi-Equilibrium Switched DAE Framework}},
  howpublished = {arXiv:2610.16768},
  year = {2026}
}

@misc{ventura2026pinn,
  author = {I. Ventura and N. Darii and P. Aristidou and M. K. Bakhshizadeh and R. Nellikkath and B. Vilmann and S. Chatzivasileiadis},
  title = {{Physics-Informed Neural Network Models for EMT Simulators}},
  howpublished = {arXiv:2511.23046},
  year = {2026}
}

@inproceedings{ruan2025ssr,
  author = {G. Ruan and M. D. Ilic and L. Xie},
  title = {{Data Center Control Against Sub-Synchronous Resonance: A Data-Driven Approach}},
  booktitle = {Proc. CIGRE US National Committee Grid of the Future Symp.},
  year = {2025},
  note = {arXiv:2511.14141}
}

@article{elsayed2026dataset,
  author = {A. A. E. Elsayed and A. A. Al-Obaidi and H. E. Z. Farag},
  title = {{Characterization of High-Resolution AI Data Center Training Workloads on Single and Multiple GPU Nodes}},
  journal = {Sci. Data},
  volume = {13},
  pages = {1268},
  year = {2026},
  doi = {10.1038/s41597-026-07496-6}
}

@misc{javadi2026review,
  author = {A. B. Javadi and H. V. Padullaparti and R. Roychowdhury and S. Maharjan and Z. Wang and K. Sreenivasachar and X. Wu},
  title = {{Large Computational Load Modeling: A Comprehensive Review of Methodologies, Applications, and Industry Practices}},
  howpublished = {SSRN preprint 7664985},
  year = {2026},
  note = {Not peer reviewed}
}

@misc{powerworld_ldev1,
  author = {{PowerWorld Corporation}},
  title = {{Load Characteristic Model: LDEV1}},
  howpublished = {Simulator 24 Transient Stability Documentation},
  year = {2025},
  note = {Added in the 22 Sep. 2025 patch; based on EPRI's positive-sequence model for aggregated EV chargers},
  url = {https://www.powerworld.com/WebHelp/Content/TransientModels_HTML/Load%20Characteristic%20LDEV1.htm}
}

@techreport{wecc_cmld_spec,
  author = {{WECC Modeling and Validation Subcommittee}},
  title = {{WECC Composite Load Model Specification}},
  institution = {Western Electricity Coordinating Council},
  year = {2021},
  month = apr,
  url = {https://www.wecc.org/sites/default/files/documents/meeting/2024/WECC%20Comp%20Load%20Model%20Specification_final.pdf}
}

@inproceedings{bindal2024npsc,
  author    = {S. Bindal and A. R. Abhyankar},
  title     = {Phase-Specific Active and Reactive Power Injection in Unbalanced Distribution Network with Different Load Models},
  booktitle = {Proc. 23rd Natl. Power Syst. Conf. (NPSC)},
  address   = {Indore, India},
  year      = {2024},
  pages     = {1--6},
  doi       = {10.1109/NPSC61626.2024.10986923}
}

@misc{nerc_mod026,
  author       = {{North American Electric Reliability Corporation}},
  title        = {Reliability Standard {MOD-026-2}: Verification and Validation of Dynamic Models and Data},
  howpublished = {NERC, Atlanta, GA, USA},
  year         = {2026},
  note         = {Approved by FERC on 19 Feb. 2026. [Online]. Available: https://www.nerc.com/globalassets/standards/reliability-standards/mod/mod-026-2.pdf}
}

\end{document}